\documentclass[a4paper,fleqn]{cas-dc}

\usepackage[numbers]{natbib}

\usepackage{amsmath}
\usepackage{amssymb}
\usepackage{amsfonts}
\usepackage{bm}
\usepackage{times,float}
\usepackage{graphicx}
\usepackage[dvipsnames,svgnames]{xcolor}
\usepackage{hyperref}
\hypersetup{colorlinks=true, linkcolor=NavyBlue, citecolor=PineGreen,urlcolor=cyan}
\usepackage{physics}
\graphicspath{{FIGURES/}}
\usepackage{subcaption}
\usepackage{cuted}

\newcommand{\partd}[3][1]{%
  \ifnum#1=1
    \frac{\partial #2}{\partial #3}%
  \else
    \frac{\partial^{#1}#2}{\partial #3^{#1}}%
  \fi
}

\def\tsc#1{\csdef{#1}{\textsc{\lowercase{#1}}\xspace}}
\tsc{WGM}
\tsc{QE}

\begin{document}
\let\WriteBookmarks\relax
\def\floatpagepagefraction{1}
\def\textpagefraction{.001}

\shorttitle{}    

\shortauthors{}  

\title [mode = title]{Exact classical emergence from high-energy quantum superpositions}  




\author[1]{Juan A. Ca\~nas}[orcid = 0009-0007-5126-7281]

\cormark[1]


\ead{juan.canas@correo.nucleares.unam.mx}


\credit{Conceptualization, Formal analysis, Methodology, Writing – original draft}

\author[1]{Daniel A. Bonilla}[orcid = 0000-0002-9719-5371]

% Footnote of the second author

% Email id of the second author
\ead{daniel.bonillam@correo.nucleares.unam.mx}


\credit{Methodology, Visualization, Writing – review \& editing}


\author[2]{J. Bernal}[orcid = 0009-0003-9327-4050]

% Footnote of the third author

% Email id of the second author
\ead{jorge.bernal@dacb.ujat.mx}


\credit{Conceptualization, Methodology, Validation}

\author[1]{A. Mart\'{i}n-Ruiz}[orcid = 0000-0001-5308-5448]

% Footnote of the fourth author

% Email id of the second author
\ead{alberto.martin@nucleares.unam.mx}


\credit{Conceptualization, Supervision, Validation, Writing – review \& editing}

\cortext[1]{Corresponding author}



\affiliation[1]{organization={Instituto de Ciencias Nucleares, Universidad Nacional Aut\'{o}noma de M\'{e}xico},
%          citysep={}, % Uncomment if no comma needed between city and postcode
            postcode={04510}, 
            state={Ciudad de M\'{e}xico},
            country={M\'exico}}

\affiliation[2]{organization={Divisi\'on Acad\'emica de Ciencias B\'asicas, Universidad Ju\'arez Aut\'onoma de Tabasco},
            city={Cunduac\'an},
%          citysep={}, % Uncomment if no comma needed between city and postcode
            postcode={86690}, 
            state={Tabasco},
            country={M\'exico}}

\begin{abstract}
We examine the correspondence principle for an equiprobable superposition of high-energy eigenstates of the infinite square well using a fully analytical Fourier-based approach. We derive a closed-form asymptotic expression for the interference terms $\rho_{\alpha}^{\text{a}}(x)$ by expanding them into a geometric series of quantum Fourier coefficients. We show these terms act as functional envelopes that do not vanish individually but become asymptotically equivalent in the large-$n$ limit. Furthermore, we prove the total probability density for a superposition of $2\Delta+1$ states converges exactly to the uniform classical distribution as $\Delta \to \infty$. Dynamically, the expectation value of position reproduces the classical triangular trajectory asymptotically. Residual quantum deviations remain confined to boundary layers whose relative width vanishes under macroscopic resolution. These results establish a rigorous asymptotic realization of the classical limit for isolated bound systems in both static and dynamical contexts.
\end{abstract}


%\item 

%\nocite{*}

% Keywords

\maketitle

% Main text

%-----------------------------------------------------------

\section{Introduction}

\noindent
The transition from the probabilistic, wave-like description of quantum mechanics to the deterministic description of classical mechanics remains a fundamental problem in modern physics. 
The correspondence principle states that quantum systems should approach classical behavior in the limit of large quantum numbers \cite{Makowski_2006, Liboff_1984}. 
For a pure eigenstate of a bound system, however, the quantum probability density (QPD) does not converge pointwise to the classical probability density (CPD), even when $n\to\infty$. 
Instead, it develops increasingly rapid oscillations around the classical distribution \cite{Robinett_1995, Robinett_1996, Doncheski_2000, Robinett_2002, Yoder2006UsingCP}. 
The classical density is then recovered only after a local averaging over lengths large compared with the quantum wavelength but small compared with the system size \cite{Liboff_1984}.

\noindent
In recent works we have developed a Fourier-based formulation of this limit at the level of probability distributions \cite{Bernal_2013, Mart_n_Ruiz_2013, Ca_as_2022, universe10090351, CMB_IJMP_2024,Hernandez_Aguilar-Gutierrez_Bernalc_2023}. 
The method expresses both the QPD and the CPD as Fourier integrals and compares their coefficients in the high-quantum-number regime. 
In this representation, the fast oscillations of the QPD are encoded in the asymptotic behavior of the quantum Fourier coefficients. 
The CPD appears as the leading large-$n$ contribution, while the remaining terms are identified as quantum corrections. 
This gives a controlled route to the classical limit without imposing an \textit{ad hoc} spatial coarse graining.

\noindent
A realistic quantum state, however, is often not a single energy eigenstate but a superposition of eigenstates. 
This issue is physically relevant because coherent superpositions have now been observed in increasingly large systems, including molecular matter-wave interference beyond 25 kDa \cite{Fein2019} and Schr\"odinger cat states in mechanical oscillators with masses up to 16 micrograms \cite{Bild2023}. 
For a state
\begin{equation}
\Psi(x,t)=\sum_n c_n \psi_n(x)e^{-iE_nt/\hbar},
\end{equation}
the probability density contains diagonal and off-diagonal contributions,
\begin{align}
\rho(x,t)
=&
\sum_n |c_n|^2 |\psi_n(x)|^2
\notag\\
&\quad+
\sum_{n\neq m}
c_n c_m^{*}
\psi_n(x)\psi_m^{*}(x)
e^{-i(E_n-E_m)t/\hbar}.
\end{align}
The second term contains the interference contributions associated with quantum coherence between different energy levels.

\noindent
The question of how, and under which conditions, these off-diagonal terms become irrelevant for macroscopic observables has a long history. 
It led von Neumann to postulate a non-unitary reduction of the state vector \cite{Neumann2018}, while Zurek proposed that the continuous interaction with an environment suppresses coherence between selected states \cite{Zurek1991}. This idea became the basis of environmental decoherence, where phase coherence is not destroyed by an extra collapse postulate but is transferred into correlations between the system and uncontrollable environmental degrees of freedom \cite{Joos2003,Schlosshauer2007,Hornberger2009}.
It has since evolved into the formalisms of \textit{einselection} and \textit{Quantum Darwinism}, in which the environment selects robust pointer states and can redundantly store information about them \cite{RevModPhys.75.715,e24111520}. 
Parallel to this approach, the role of thermalization has also been debated; recent work argues that thermalization alone cannot guarantee classicality without external amplification processes \cite{TIRANDAZ20191677}.

\noindent
For isolated bound systems, where environmental monitoring is absent, the persistence of interference terms in the high-energy regime has been used to question the validity of the correspondence principle \cite{Rosen1964, Home1984, Kmmel1955, PhysRevA.36.2995}. 
Related concerns also appear in semiclassical WKB-type approaches, which describe phase information efficiently but do not by themselves remove off-diagonal contributions in position-space densities or observables \cite{Korsch1978}. 
In particular, Cabrera and Kiwi studied superpositions of a small number of high-energy harmonic oscillator states and concluded that ``the correspondence principle has only heuristic value: quantum and classical mechanics only partially overlap in the macroscopic domain'' \cite{PhysRevA.36.2995}.

\noindent
In this work we address this problem for an equiprobable superposition of high-energy eigenstates of the infinite square well (ISW). 
The ISW is analytically tractable and provides a simple setting where confinement, interference, and the classical limit can be studied exactly. 
It also has subtle mathematical features related to its rigid boundaries. 
For example, the singular nature of the potential leads to the known breakdown of Ehrenfest's theorem when the standard momentum operator is used. 
Recent studies have restored the theorem by introducing strictly self-adjoint momentum operators \cite{ALBRECHT2023169289, 10.1063/5.0178419}. 
Although our analysis does not rely on Ehrenfest's theorem, we show that the expectation value of position reproduces the classical triangular trajectory of a particle bouncing between rigid walls.

\noindent
Another common route to the quantum-classical transition is the Wigner-function formalism. 
For systems with rigid boundaries, however, this approach is technically delicate because the singular walls generate nonlocal interference structures and rapidly oscillating phase-space contributions \cite{BELLONI201425, 10.1063/1.1504885}. 
We therefore use a complementary configuration-space approach and analyze the classical limit directly through the spatial probability density.

\noindent
By extending the Fourier-based asymptotic method to interference terms, we derive closed expressions for their high-energy contribution to the probability density. 
We show that these terms do not vanish individually. 
Instead, in the large-$n$ limit, they form smooth envelope functions that modulate the rapidly oscillating quantum density. 
When the superposition contains a sufficiently large set of neighboring high-energy states, representing the finite energy resolution of macroscopic measurements, the full QPD converges to the uniform CPD expected for a classical particle in the ISW. 
The remaining quantum distortions are confined to narrow boundary regions and become negligible at macroscopic observational scales.

\noindent
Our results resolve the apparent conflict raised by Cabrera and Kiwi. 
The persistence of interference terms does not invalidate the correspondence principle. 
Rather, when high-energy superpositions are treated with a controlled asymptotic expansion and finite macroscopic resolution is included, the QPD approaches its classical counterpart in a precise and quantitative way. 
In this sense, the correspondence principle is not only a heuristic rule but an asymptotic property of probability densities in the macroscopic limit.

\noindent
The paper is organized as follows. 
Section \ref{Sec_formalism} summarizes the Fourier-based formulation of the correspondence principle. 
Section \ref{Sec:ISW} applies it to the infinite square well for pure eigenstates and transition terms. 
Section \ref{SEC:Interference} introduces the equiprobable high-energy superposition. 
Section \ref{SEC:PD} gives the asymptotic analysis of its probability density and proves its convergence to the classical limit. 
Section \ref{Sec:Time} studies the time-dependent expectation value of position and compares it with the classical trajectory. 
Finally, Section \ref{Sec:Conclusion} presents the conclusions.

\section{  Fourier-based asymptotic framework for the classical limit }
\label{Sec_formalism}

In this section we summarize the analytical framework developed in Refs.~\cite{Bernal_2013,Mart_n_Ruiz_2013,Ca_as_2022, universe10090351, CMB_IJMP_2024}, which provides a systematic formulation of the correspondence principle based on the Fourier representation of probability densities. The essential elements of the method are presented here in a concise form in order to keep the discussion self-contained, while the full technical derivations can be found in the cited works. This formalism allows one to analyze the classical limit by studying the asymptotic behavior of the Fourier coefficients of the QPD, which can then be directly compared with those of the corresponding CPD. To facilitate reading and avoid any ambiguity, let us clarify in advance the notation used for the different quantities throughout this manuscript: the subscript ``cl'' indicates classical quantities, such as the CPD ($\rho_{\text{cl}}$); the superscript ``qm'' stands for quantities derived from the quantum formalism, such as the QPD ($\rho^{\text{qm}}$); and the superscript ``a'' specifically denotes quantities arising from the asymptotic approximation in the high-energy limit.

Although classical mechanics is fundamentally deterministic, a probability distribution naturally arises when the system is sampled at an arbitrary time. For instance, consider a particle undergoing periodic motion between $x_a$ and $x_b$ with period $\tau$. If the observation time is assumed to be uniformly distributed over one period, the CPD $\rho_{\mathrm{cl}}(x)$ is defined so that the probability of finding the particle in the infinitesimal interval $[x,x+dx]$ is
\begin{align}
    dP_{\mathrm{cl}} =\rho_{\mathrm{cl}}(x)\,dx .
\end{align}
This probability is proportional to the fraction of time that the particle spends within this interval during its periodic motion. Since the particle crosses each position twice during a full cycle (once in each direction), the probability can be written as
\begin{align}
    dP_{\mathrm{cl}} =\frac{2\,dt(x)}{\tau}.
\end{align}
Using $dt = dx/|v(x)|$, where $v(x)$ is the local speed, one obtains the CPD
\begin{equation}
\rho_{\mathrm{cl}}(x)=\frac{2}{\tau}\frac{dt(x)}{dx}
=\frac{2}{\tau |v(x)|},
\label{CPD_DEF}
\end{equation}
which reflects the fact that the particle is more likely to be observed in regions where its velocity is smaller, i.e., near the classical turning points. This probability distribution provides the natural quantity to be compared with the QPD in the correspondence limit.

In quantum mechanics, the probability density arises directly from the Born rule. Given the wave function of the system $\psi(x)$, the QPD is given by
\begin{equation}
\rho^{\mathrm{qm}}(x)=|\psi(x)|^2 .
\end{equation}
In contrast with the classical case, the probabilistic interpretation does not originate from an external sampling procedure but is an intrinsic feature of the quantum state itself.

When $\psi_n(x)$ represents an energy eigenstate with quantum number $n$, a connection between the classical and quantum descriptions is expected to emerge in the high-quantum-number regime $n\gg1$, in accordance with the correspondence principle. Comparisons between the two distributions show that the CPD can be interpreted as a locally averaged version of the rapidly oscillatory QPD
\cite{Robinett_1995, Robinett_1996, Doncheski_2000, Robinett_2002, Yoder2006UsingCP}. However, the QPD $\rho_n^{\mathrm{qm}}(x)$ typically exhibits rapidly oscillatory behavior for large quantum numbers and therefore does not converge pointwise to the smooth CPD $\rho_{\mathrm{cl}}(x)$. Instead, Bohr’s correspondence principle must be interpreted in a coarse-grained or distributional sense, reflecting the fact that macroscopic measurements possess finite spatial resolution. In this interpretation the classical distribution emerges when the QPD is locally averaged over a small interval $\epsilon_n$
\cite{Liboff_1984}
\begin{equation}
\rho_{\mathrm{cl}} (x)
=
\lim_{n\to\infty} \; 
\frac{1}{2\epsilon_n} \;
\int_{x-\epsilon_n}^{x+\epsilon_n}
\rho^{\mathrm{qm}}_{n}(y)\,dy ,
\label{LocalAverage}
\end{equation}
where the averaging scale $\epsilon_n$ is assumed to be large compared with the wavelength of the quantum oscillations but small compared with the macroscopic length scale of the system.

With this interpretation in mind, the correspondence principle can be implemented directly at the level of the spectral representation of the probability density. In the formalism developed in Refs.~\cite{Bernal_2013,Mart_n_Ruiz_2013,Ca_as_2022, universe10090351, CMB_IJMP_2024}, the classical limit is analyzed by expressing both the quantum and classical probability densities through their Fourier representations and comparing the behavior of their corresponding coefficients in the large-quantum-number regime. Specifically, the probability density can be written as
\begin{equation}
\rho(x)
=
\frac{1}{2\pi\hbar}
\int f(p)\,e^{ipx/\hbar}\,dp ,
\end{equation}
where $f(p)$ denotes the Fourier coefficient associated with the distribution.

Within this representation the rapidly oscillatory structure of the quantum density is encoded in the coefficients $f^{\mathrm{qm}}_{n}(p)$. 
The correspondence principle is then implemented by studying the asymptotic behavior of these coefficients for large quantum numbers and showing that they approach the classical coefficients $f^{\mathrm{cl}}(p)$:
\begin{equation}
f^{\mathrm{qm}}_{n}(p)
\sim
f_{\mathrm{cl}}(p)
+
\mathcal{O}(1/n).
\end{equation}
Consequently, the CPD emerges as the leading asymptotic contribution of the quantum expansion, while the subleading terms encode residual quantum corrections that become progressively suppressed in the macroscopic limit.

At this point it is useful to clarify the connection between the Fourier representation and the coarse-grained interpretation of the classical limit introduced in Eq.~(\ref{LocalAverage}). Substituting the Fourier expansion of the quantum probability density into the local average, one finds that the averaging procedure produces a multiplicative factor of the form
\begin{align}
    \frac{1}{2\epsilon_n}
\int_{x-\epsilon_n}^{x+\epsilon_n}
e^{ipy/\hbar}\,dy
=
e^{ipx/\hbar} \; 
\frac{\sin(p\epsilon_n/\hbar)}{p\epsilon_n/\hbar},
\end{align}
so that each Fourier component is weighted by a $\operatorname{sinc}$ function. This factor acts as a low-pass filter in momentum space: Fourier components satisfying $p\epsilon_n/\hbar \ll 1$ are essentially unaffected, while higher-momentum components are suppressed.

From this perspective, spatial coarse graining over a length $\epsilon_n$ is equivalent to restricting the relevant Fourier modes to a range $|p| \lesssim p_{\mathrm{cg}}$, with
\begin{align}
    p_{\mathrm{cg}} \, \epsilon_n/\hbar \sim \mathcal{O}(1) ,
\end{align}
up to numerical factors depending on the precise definition of the cutoff. Conversely, when the asymptotic expansion of the Fourier coefficients is valid only within a finite domain $|p| \lesssim p_{\max}$, this restriction implies an effective spatial resolution
\begin{align}
    \Delta x \sim \frac{\hbar}{p_{\max}}.
\end{align}
Therefore, the limitation on the Fourier variable should be understood as the Fourier-space counterpart of spatial coarse graining, rather than as an independent assumption.

So far, this formalism has been applied primarily to individual eigenstates of bound quantum systems. In realistic situations, however, the quantum state is generally described by a superposition of energy eigenstates. When the associated probability density is computed, cross terms between different eigenfunctions appear, giving rise to interference contributions that may introduce non-classical features.

This issue was emphasized by Cabrera and Kiwi \cite{PhysRevA.36.2995}, who analyzed superpositions of a small number of harmonic oscillator eigenstates with large quantum numbers. Their results suggested that significant interference effects may persist even in the large-$n$ regime, leading them to question the validity of the correspondence principle in this context.

In the next section we extend the present formalism to the case of an equiprobable superposition of eigenstates in the ISW. Despite its apparent simplicity, the ISW provides an analytically controlled setting in which interference and confinement can be treated exactly. The analysis of this model allows us to determine explicitly how the interference contributions behave in the high-energy regime and to clarify the role of the correspondence principle for superposed quantum states.

\

\section{Classical limit of diagonal and interference terms in the infinite square well} \label{Sec:ISW}

As a paradigmatic example of a confined quantum system, we consider a particle of mass $m$ confined within a one-dimensional box of length $L$ with rigid boundaries. For this system the classical motion corresponds to a particle moving with constant speed between perfectly reflecting walls. 
Using Eq.~\eqref{CPD_DEF}, and noting that the velocity remains constant throughout the motion, the period of the classical trajectory is $\tau=2L/v$. The resulting CPD is therefore uniform within the accessible region:
\begin{equation}
\rho_{\mathrm{cl}}(x)=\frac{1}{L}H(x)H(L-x),
\label{CPD_ISW}
\end{equation}
where $H(x)$ denotes the Heaviside step function enforcing the spatial confinement.

The stationary quantum states follow from the Schr\"odinger equation with Dirichlet boundary conditions at the walls. The normalized eigenfunctions and eigenenergies are
\begin{align}
\psi_n(x) &= \sqrt{\frac{2}{L}}
\sin\!\left(\frac{n\pi x}{L}\right)H(x)H(L-x),
\label{WaveFunct}\\[6pt]
E_n &= \frac{p_n^2}{2m}
= \frac{n^2\pi^2\hbar^2}{2mL^2},
\label{Eigenfun}
\end{align}
where $n=1,2,\dots$ labels the energy eigenstates. The corresponding QPD $\rho_n^{\mathrm{qm}}(x)=|\psi_n(x)|^2$ displays a sequence of oscillatory nodes and antinodes across the well. As the quantum number increases, the number of oscillations grows and the density becomes increasingly structured on short spatial scales. However, when averaged over intervals larger than the oscillation wavelength, the distribution approaches the classical uniform value $1/L$, in accordance with the correspondence principle.

We now apply the Fourier-based asymptotic formalism introduced in Sec.~\ref{Sec_formalism} to the ISW.
By applying the Fourier transform to $\rho_n^{\mathrm{qm}}(x)$, we obtain the quantum Fourier coefficients
\begin{equation}
\rho ^{\mathrm{qm}} _{n} (p)  = \frac{i}{2\pi \hbar Q} \, \frac{ 1- e ^{ - iQ  } }{  1 - \big( \frac{ Q}{2 \pi n } \big) ^{2}  },
\label{Fourier_Eigenstate}
\end{equation}
where $Q \equiv pL/\hbar$. In the high-energy regime $n\gg1$, the denominator can be expanded as a geometric series in powers of $1/n$, provided that $|p|<2p_n$, where $p_n=n\pi\hbar/L$ is the quantum momentum associated with the $n$th eigenstate. This restriction has a clear physical interpretation: it corresponds precisely to the classical momentum domain accessible to a particle with energy $E_n$, reflecting the bounded classical phase space of the system \cite{Bernal_2013}.

Retaining the leading term of this expansion and performing the inverse Fourier transform yields the asymptotic form of the QPD in coordinate space:
\begin{equation}
\rho_n^{\text{a}}(x) \sim \frac{1}{L \pi}
\left[
\text{Si}(2n\pi(1-\chi)) + \text{Si}(2n\pi\chi)
\right]
+ \mathcal{O}(1/n^2),
\end{equation}
where $\chi = x/L$ and $\text{Si}(z)$ is the sine integral \cite{Gradshteyn2015}.

\begin{figure}
\vspace{.4cm}
\includegraphics[scale=0.23]{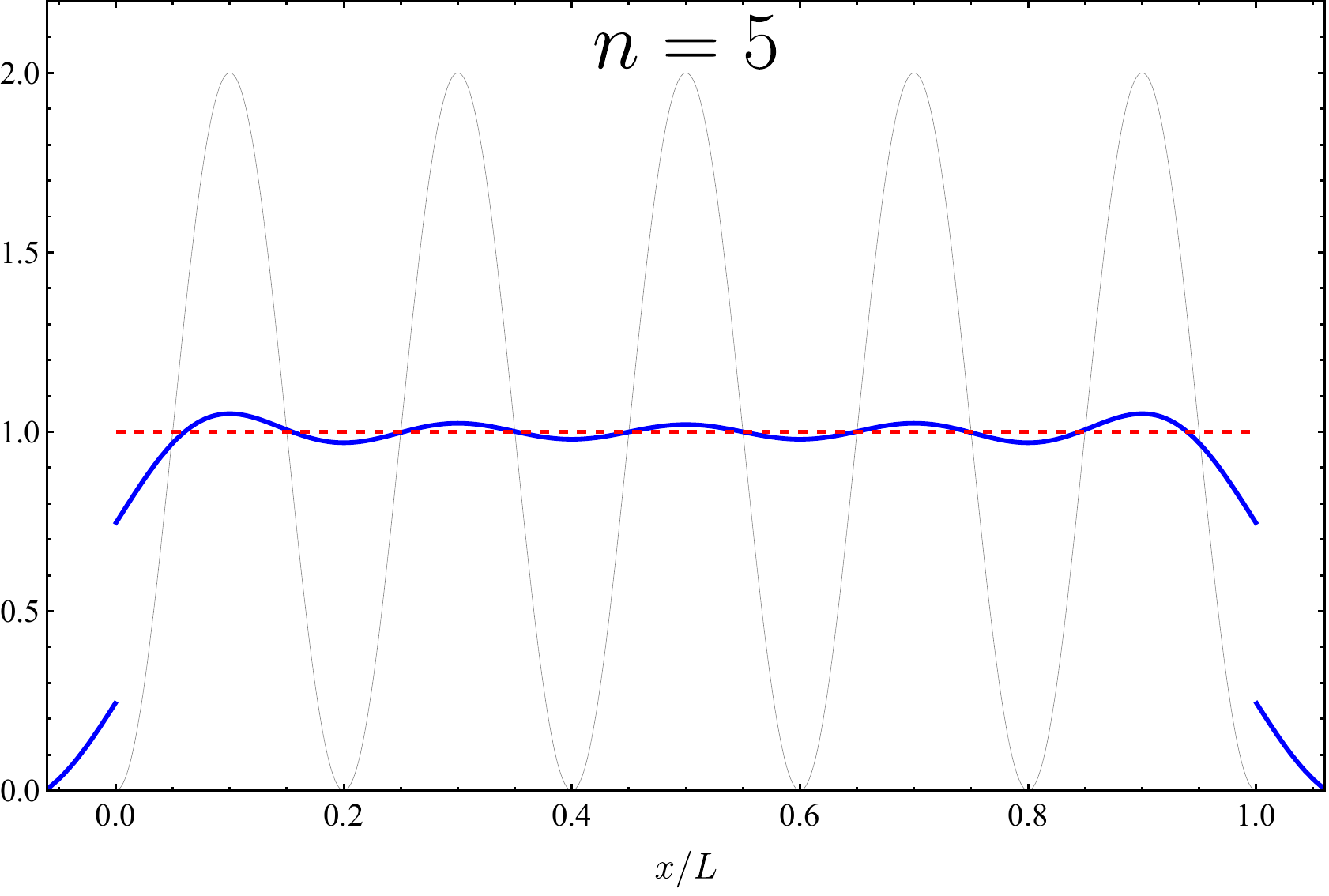} \hspace{.5cm}
\includegraphics[scale=0.23]{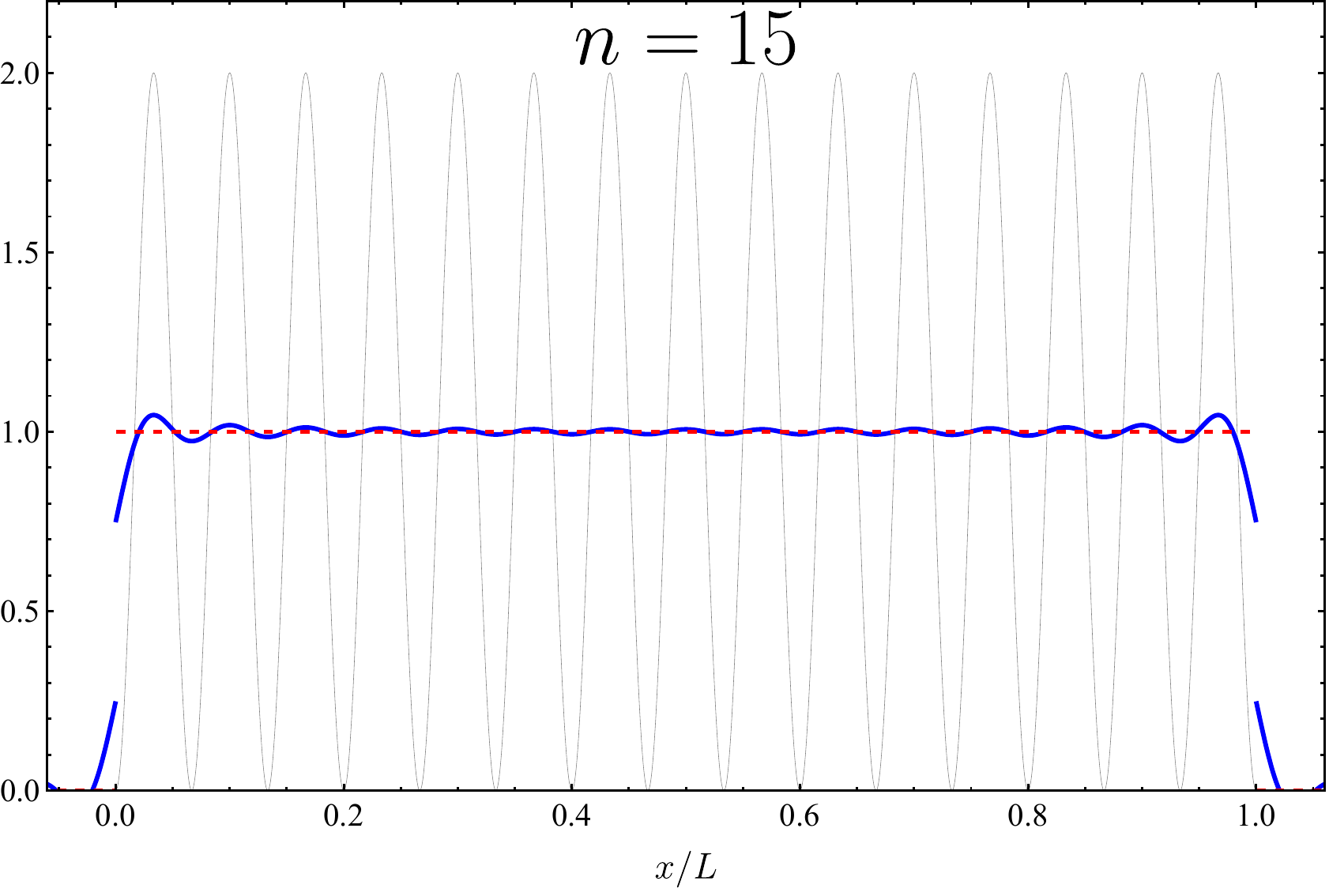} 
\caption{Dimensionless plots of QPD $ L \rho ^{\mathrm{qm}}_{n} (x) $ (continuous gray), asymptotic distribution $ L \rho ^{\mathrm{a}}_{n} (x)$ (continous blue) and CPD $ L \rho _{\mathrm{cl}} (x)$ (dashed red) for the quantum numbers $n=5$ (top) and $n=15$ (bottom). As discussed in \cite{universe10090351},  both the small-scale quantum oscillations and the anomalous behavior near the box walls arise from the relatively low values of $n$ considered here. In a truly classical regime, these features would be strongly suppressed.}
\label{fig:SquareWell}
\end{figure}

In the formal limit $n\to\infty$, the sine-integral functions approach step functions, thereby reproducing exactly the Heaviside structure of the classical density $\rho_{\text{cl}}(x)$. 
However, the subleading $\mathcal{O}(1/n^2)$ contributions contain rapidly oscillatory terms that encode residual quantum structure in the high-energy regime. Although these oscillations become progressively suppressed at macroscopic scales, they remain an intrinsic signature of the quantum-to-classical transition. A detailed analysis of these oscillations for both the ISW and the quantum bouncer can be found at \cite{universe10090351}. A comparison between classical, quantum and asymptotic distributions is shown in Fig.~\ref{fig:SquareWell}.

We now analyze the asymptotic behavior of the interference (off-diagonal) contributions of the form
\begin{equation}
\rho_{\alpha}^{\mathrm{qm}}(x)
=
\frac{2 \sin \left(\frac{\pi x n}{L}\right)
\sin \left(\frac{\pi x (n+\alpha)}{L}\right)}{L}.
\end{equation}
Unlike the diagonal terms associated with individual eigenstates, these interference contributions do not possess a direct classical counterpart. For this reason it is often assumed that it must vanish in the classical limit in order for the correspondence principle to hold. Indeed, several authors have argued that the persistence of such interference terms at large quantum numbers challenges the validity of the correspondence principle
\cite{Rosen1964,Home1984,PhysRevA.36.2995,Kmmel1955}.

The Fourier transform of the interference term is given by
\begin{equation}
\rho_{\alpha} ^{\mathrm{qm}} (p)\!
=
\!\frac{n (n+\alpha)}{(n+\alpha/2)^2}
\frac{ i Q \left[ 1-(-1)^{\alpha} e^{-i Q} \right] }
{2\pi \hbar  \left(Q^2 - \pi^2 \alpha^2\right)
\left[ 1 - \left(\frac{Q}{\pi(2n + \alpha )}\right)^2 \right]}, \label{fourier_exp_nondiagonal}
\end{equation}
where $Q$ is defined as before. As a consistency check, setting $\alpha=0$ reduces this expression to the Fourier transform of the diagonal contribution, reproducing Eq.~\eqref{Fourier_Eigenstate}. The prefactor
\begin{align}
    \Gamma_{\alpha}^{n} =\frac{n(n+\alpha)}{(n+\alpha/2)^2}
\end{align}
approaches unity in the high-energy limit $n\gg1$.

For large quantum numbers, the denominator in Eq.~(\ref{fourier_exp_nondiagonal}) can be expanded as a geometric series in powers of $1/n$, yielding
\begin{equation}
\rho _{\alpha} ^{\mathrm{qm}} (p) = \frac{\Gamma ^{n} _{\alpha} i Q \left[ 1 - ( - 1 ) ^{\alpha} e ^{- i Q} \right] }{2 \pi \hbar  \left( Q ^{2} - \pi ^{2} \alpha ^{2} \right) } \sum _{k = 0} ^{\infty}
\left( \frac{Q}{\pi(2n + \alpha )}\right)^{2 k },
\label{Trans_AsympFourier}
\end{equation}
which provides the starting point for the asymptotic analysis of the interference contribution in the large-$n$ regime. The geometric expansion is valid provided that $|Q|<\pi(2n+\alpha)$. Beyond its mathematical origin, this condition also carries a clear physical interpretation: it restricts the momentum support of the Fourier coefficients to a finite interval associated with the momenta of the two states involved in the interference term. In particular, the characteristic scale is set by the average quantum number $n+\alpha/2$, reflecting the effective momentum associated with the pair of levels $(n,n+\alpha)$.

Retaining only the leading term of the expansion and performing the inverse Fourier transform, we obtain the asymptotic expression for the interference contribution:
\begin{align}
    &\rho_{\alpha} ^{\text{a}}(x)  =  \frac{i \Gamma^{N}_{\alpha}}{4 \pi L}\Big\{ e^{-\frac{i \pi  \alpha  x}{L}} \Big[\text{Ei}\left(-\zeta_{\alpha}^{1}(x)\right)-\text{Ei}\left(\zeta_{0}^{1}(x)\right)\notag \\[6pt]
    & \quad +\text{Ei}\left(-\zeta_{0}^{0}(x)\right) - \text{Ei}\left(\zeta_{\alpha}^{0}(x)\right) \big] +e^{\frac{i \pi  \alpha  x}{L}} \left[-\text{Ei}\left(\zeta_{\alpha}^{1}(x)\right)\right.  \notag \\[6pt]
    & \quad \left. +\text{Ei}\left(-\zeta_{\alpha}^{0}(x)\right)-\text{Ei}\left(\zeta_{0}^{0}(x)\right)+\text{Ei}\left(-\zeta_{0}^{1}(x)\right)\right]\Big\},
    \label{Asymp_NDTerm}
\end{align}
where $\zeta_{\alpha}^{\xi}(x) = 2 i \pi (x)^{1-\xi} (L-x)^{\xi} (n+ \alpha)/L$, 
and $\text{Ei}(z)$ denotes the exponential integral function \cite{Gradshteyn2015}.

\begin{figure}
\vspace{.4cm}
\includegraphics[scale=0.23]{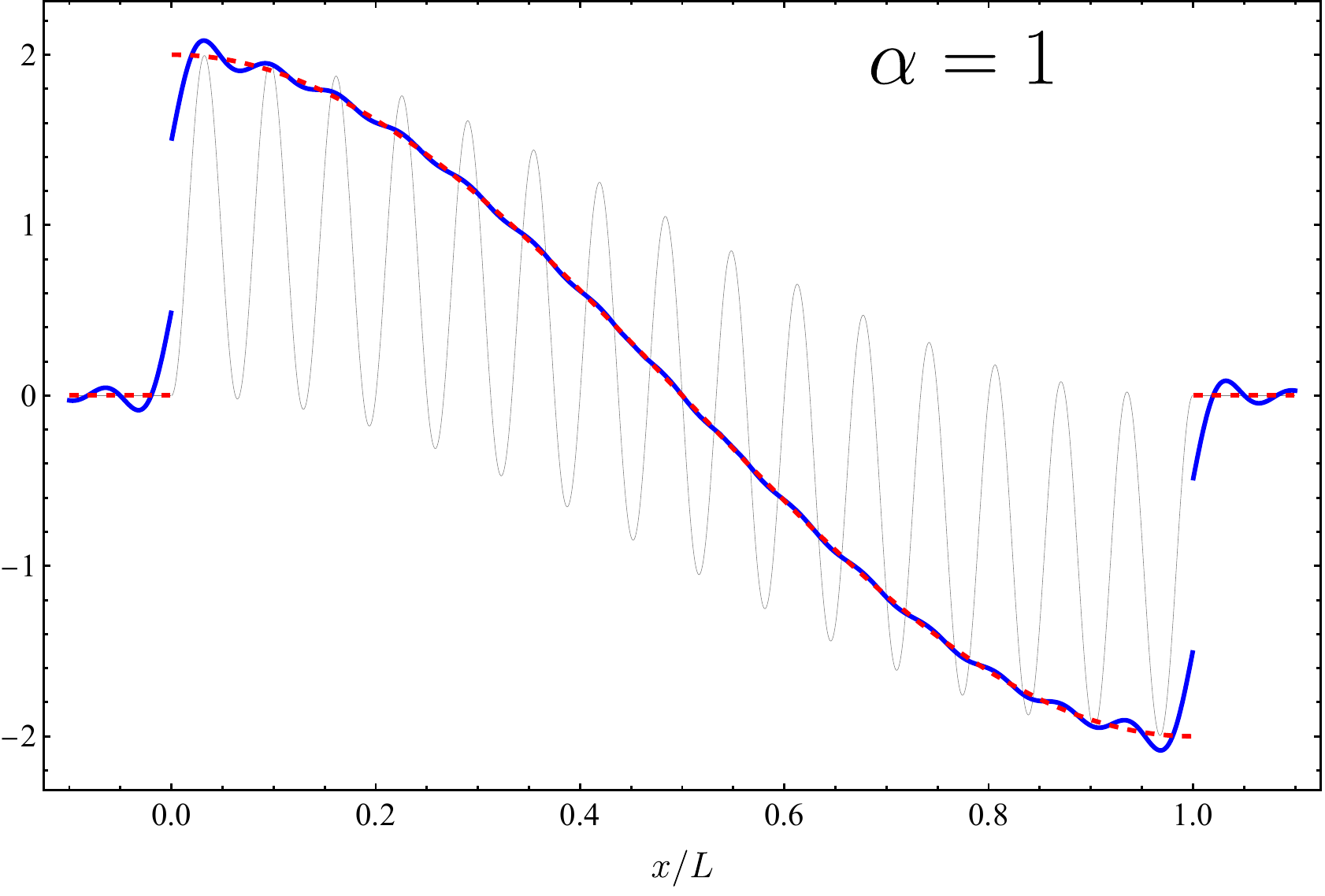} \hspace{.5cm}
\includegraphics[scale=0.23]{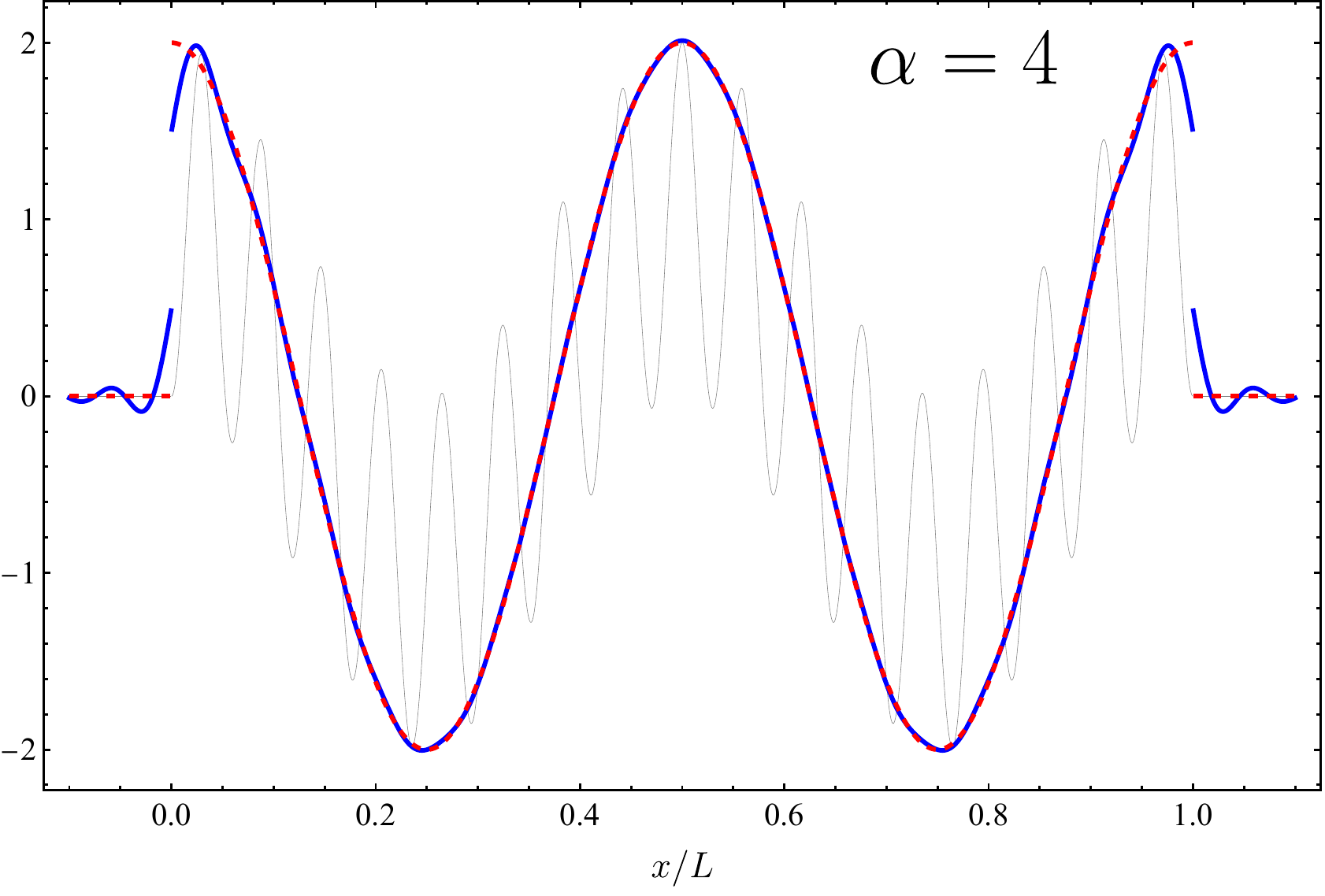} 
\caption{Plots of the interference term $L \rho^{\mathrm{qm}}_\alpha(x)$ (gray), its large-$n$ asymptotic behavior $L \rho_{\alpha} ^{\text{a}}(x)$ (blue), and the $n\to\infty$ limit from eq. \eqref{Asymp_NDTerm_inf} (red dashed), for $\alpha=1$ (top) and $\alpha=4$ (bottom), $n=15$ in both cases. Unlike the case of pure eigenstates, the asymptotic behavior does not represent a local average of the quantum oscillations; instead, it functions as a functional envelope for the interference terms.}\label{AsympNDTerms}
\end{figure}

To analyze the classical limit, we examine the behavior of $\rho_{\alpha}^{\text{a}}(x)$ as $n \to \infty$. Because the arguments $\zeta_{\alpha}^{\xi}(x)$ are purely imaginary and scale linearly with $n$, they become arbitrarily large in the high-energy regime. Using the relation $\text{Ei}(iy) = \text{Ci} (y) + i \text{Si}(y)$ and the formal asymptotic behavior of the sine integral, $\text{Si} (y) \sim \frac{\pi}{2} \text{sgn}(y) - \frac{\cos y}{y}$, the leading-order terms in Eq.~\eqref{Asymp_NDTerm} reorganize into the Heaviside structure $H(x)H(L-x)$. This ensures that the interference contribution is strictly confined to the classically allowed region $x \in [0,L]$, while the remaining terms of order $\mathcal{O}(1/n)$ encode the residual quantum oscillations. Taking this limit yields the simplified asymptotic expression: 
\begin{equation}
\rho_{\alpha} ^{\text{a}}(x) 
\approx 
\frac{2}{L}
\cos\!\left(\frac{\pi \alpha x}{L}\right)
H(x)H(L-x).
\label{Asymp_NDTerm_inf}
\end{equation}
Figure~\ref{AsympNDTerms} compares the exact quantum interference terms with their asymptotic form for different values of $n$. The figure clearly shows that the interference contributions do not vanish in the large-$n$ limit. Instead, they approach the bounded oscillatory envelope given by Eq.~\eqref{Asymp_NDTerm_inf}.

The persistence of these interference contributions does not signal a breakdown of the correspondence principle. Rather, it reflects the spectral structure of the system in the high-energy regime. Indeed, for large quantum numbers the eigenstates $\psi_n$ and $\psi_{n+\alpha}$ become progressively closer in energy and structure. Since the spectrum of the ISW grows quadratically with $n$, the energy separation between nearby levels behaves as $|E_n - E_{n+\alpha}|\approx \frac{\pi^2 \hbar^2}{mL^2} n |\alpha|$ which implies the relative spacing
\begin{equation}
\frac{|E_n - E_{n+\alpha}|}{E_n}
\approx
2\frac{|\alpha|}{n}.
\label{Diff_energies}
\end{equation}
Therefore, whenever $\alpha\ll n$, the two levels become effectively quasidegenerate. Henceforth, in this regime the interference between the corresponding eigenstates does not involve a significant change in energy and thus remains dynamically relevant even in the high-energy limit.

Equation~\eqref{Diff_energies} also shows that, for sufficiently large $n$, the condition $\alpha\ll n$ can still be satisfied even when $\alpha$ itself is large. Consequently, a broad range of interference terms remains effectively quasidegenerate in the macroscopic limit, extending well beyond the case of transitions between immediately neighboring levels.

\section{Equiprobable superpositions in the high-energy regime}
\label{SEC:Interference}

We now extend the analysis to superpositions of eigenstates of the ISW. In contrast with the conventional construction of wave packets strongly localized around a single energy level (typically described by Gaussian weight distributions) we consider a different type of superposition appropriate to the macroscopic regime. This choice is motivated by the physical interpretation of superpositions at very large quantum numbers.

Consider a system prepared in a superposition of eigenstates with energies $E_{n+k}$, where $k=-\Delta,-\Delta+1,\ldots,\Delta$, and $n\gg1$. Within the standard interpretation of quantum mechanics, such a state represents a situation in which an energy measurement yields a value within the interval $E_n\pm\mathcal{E}_\Delta$, where $E_n$ denotes the mean energy of the superposition. The measurement uncertainty $\mathcal{E}_\Delta$ vanishes only when the system occupies a single eigenstate, in which case the energy is sharply defined as $E_n$.

To estimate the typical scales involved, consider a thermal neutron with energy $E\sim10^{-21}$ J confined in a box of length $L=1$ mm. Using the energy spectrum of the ISW, this corresponds to quantum numbers of order $n\sim10^7$. If the energy measurement has a relative uncertainty of about $1\%$, the maximal deviation from the mean energy satisfies
\begin{equation}
|E_{n}-E_{n+\Delta}|
\approx
10^{-23}\text{ J}
\quad\Rightarrow\quad
\Delta\sim10^5.
\end{equation}
This estimate shows that the range parameter $\Delta$ can become very large while still corresponding to a small relative uncertainty in the measured energy. Furthermore, as discussed in the previous section, eigenstates in the regime $n\gg1$ become effectively quasidegenerate. Since their energies differ only by a very small fraction of the mean energy, there is no physical reason to favor one state over another within the allowed energy window. It is therefore natural to model this high-energy superposition as an equiprobable distribution of states, leading to the wave function
\begin{equation}
\Psi(x)=\frac{1}{\sqrt{2\Delta+1}}
\sum_{k=-\Delta}^{\Delta}\psi_{n+k}(x),
\label{wave_Superp}
\end{equation}
which will serve as the starting point of our analysis.

It is worth pointing out that the equiprobable superposition should not be understood as the only physically meaningful choice. 
One expects that any sufficiently broad distribution of eigenstates, such as a Gaussian or Poisson-like distribution centered around a large quantum number, should lead to the same classical limit obtained in the present case.

\section{Macroscopic Probability Density of the Superposed State}
\label{SEC:PD}

We now examine the probability density $|\Psi(x)|^2$ associated with the superposition of states in the high-energy limit. Expanding the square in Eq.~\eqref{wave_Superp} separates the diagonal contributions, which reproduce the classical density $\rho_{\mathrm{cl}}$, from the off-diagonal interference terms. Grouping the latter according to the index difference $\alpha$ yields
\begin{equation}
\rho(x)
=
\left| \, 
\rho_{\mathrm{cl}}
+
\frac{2}{2\Delta+1}
\sum_{\alpha=1}^{2\Delta} \, 
(2\Delta-\alpha+1) \, 
\rho_{\alpha}^{\text{a}}(x) \, 
\right|.
\label{QPD_Nosum}
\end{equation}

Using the asymptotic expression for the interference contributions $\rho_{\alpha}^{\text{a}}(x)$ given in Eq.~\eqref{Asymp_NDTerm_inf}, the summation appearing in the second term of Eq.~\eqref{QPD_Nosum} can be evaluated analytically. For completeness, the intermediate algebraic steps are presented in Appendix~\ref{Ap:A}.

Introducing the variables $N=2\Delta$ and $y=\pi x/L$ in the sum $S$ defined in Eq.~\eqref{ap:sum}, and substituting the resulting expression into Eq.~\eqref{QPD_Nosum}, we obtain the following closed-form expression for the probability density:
\begin{align}
\rho(x) =& \frac{H(x)H(L-x)}{L}
\left| \vphantom{\frac{\csc^2\left(\frac{\pi x}{2L}\right)
\left[\cos\left(\frac{\pi x}{L}\right)-\cos\left(\frac{\pi(2\Delta+1)x}{L}\right)\right]}
{2\Delta+1}}
1-\frac{4\Delta}{2\Delta+1} \right. \notag\\[6pt]
&\left. + \frac{\csc^2\left(\frac{\pi x}{2L}\right)
\left[\cos\left(\frac{\pi x}{L}\right)-\cos\left(\frac{\pi(2\Delta+1)x}{L}\right)\right]}
{2\Delta+1} \right|.
\label{PD_Final}
\end{align}
%%La versión con wide text está acá comentada
%\begin{widetext}
%    \begin{equation}
%    \rho(x)= \frac{1}{L}\left| 1 +\frac{\csc ^2\left(\frac{\pi  x}{2 L}\right) \left(\cos \left(\frac{\pi  x}{L}\right)-\cos \left(\frac{\pi  (2 \Delta +1) x}{L}\right)\right)-4 \Delta }{2 \Delta +1} \right|H (x) H (L-x).
%    \label{PD_Final}
%\end{equation}
%\end{widetext}

\begin{figure}
\vspace{.4cm}
\includegraphics[scale=0.23]{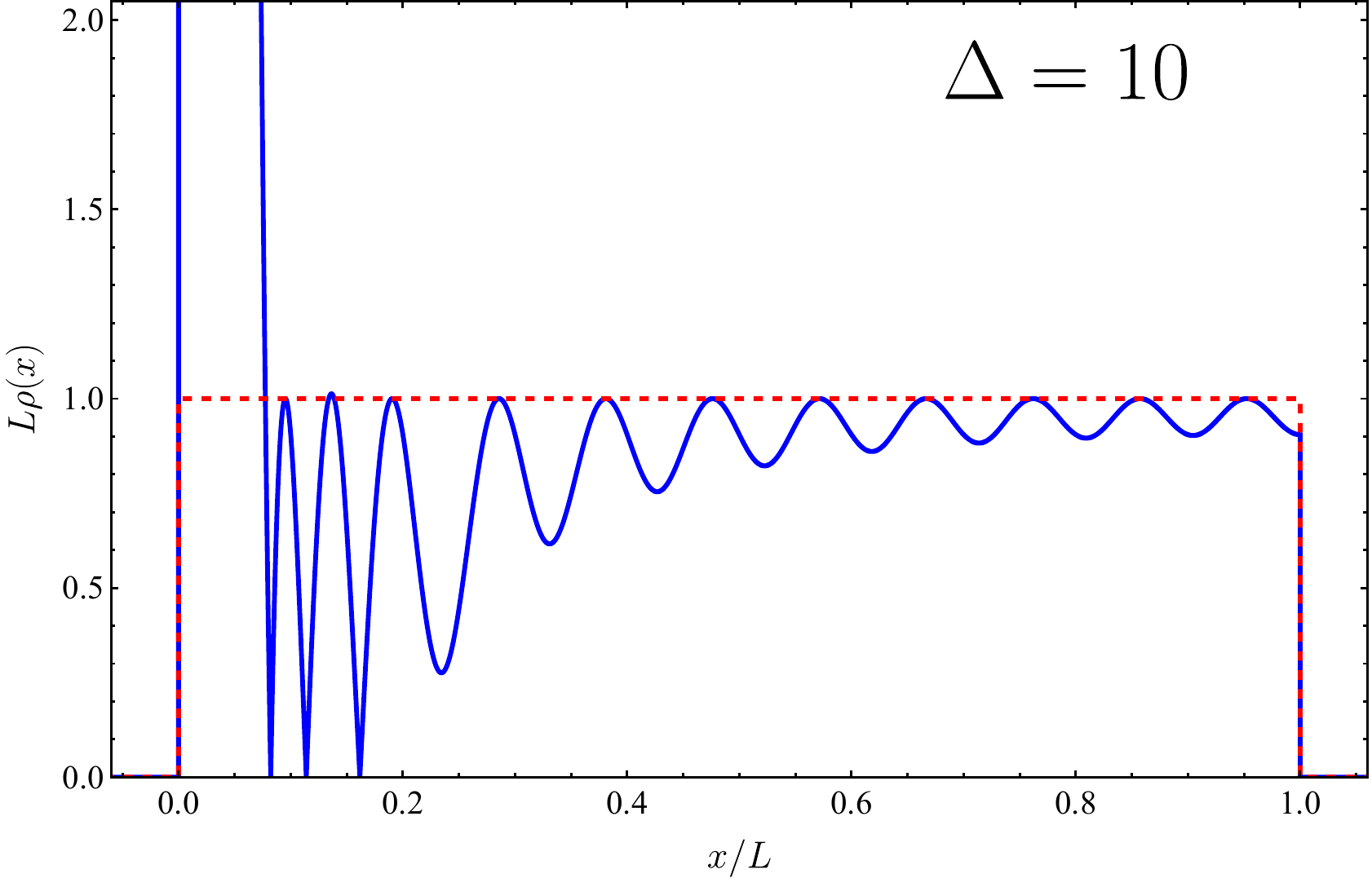}

\includegraphics[scale=0.23]{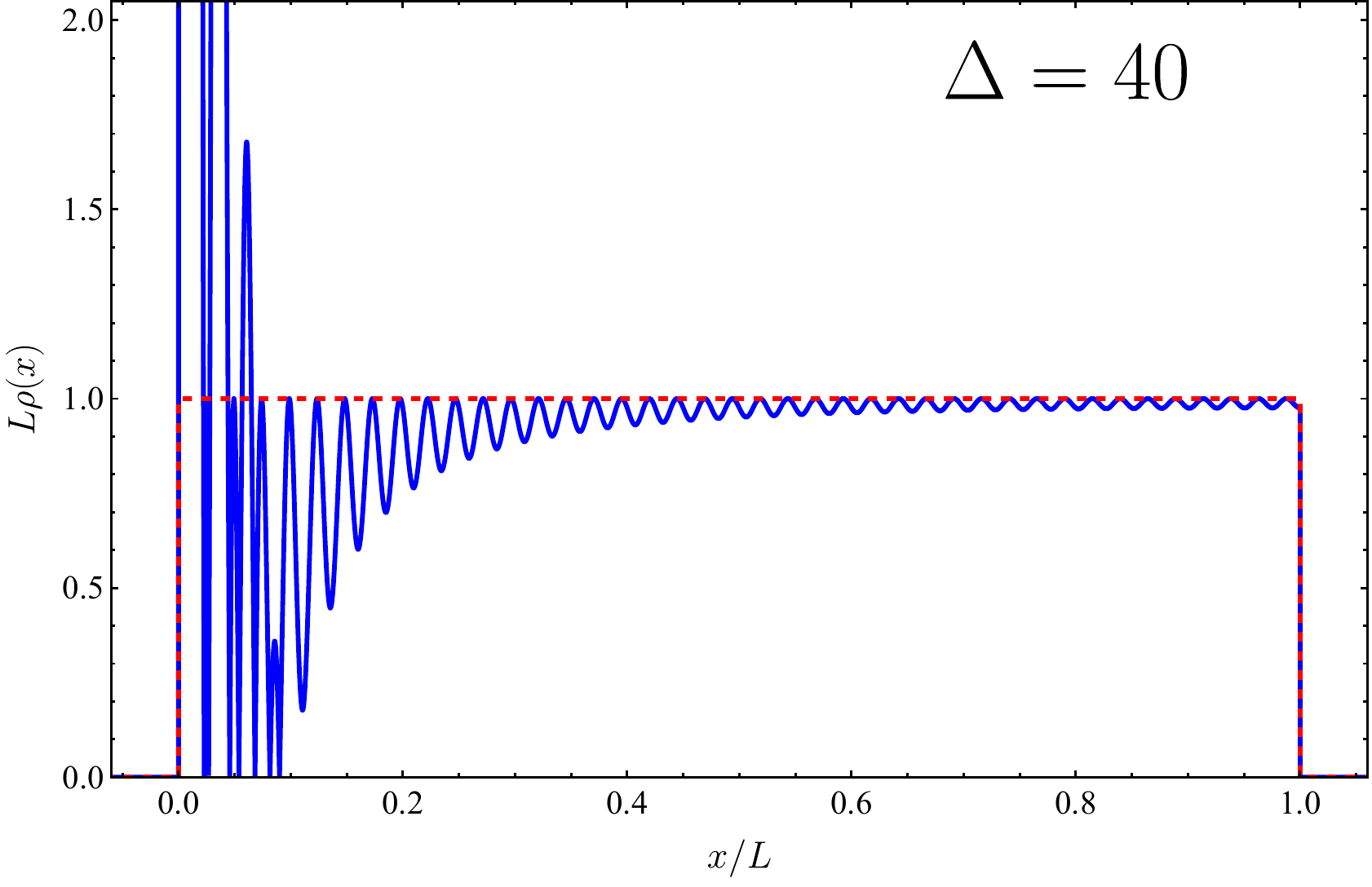} 

\includegraphics[scale=0.23]{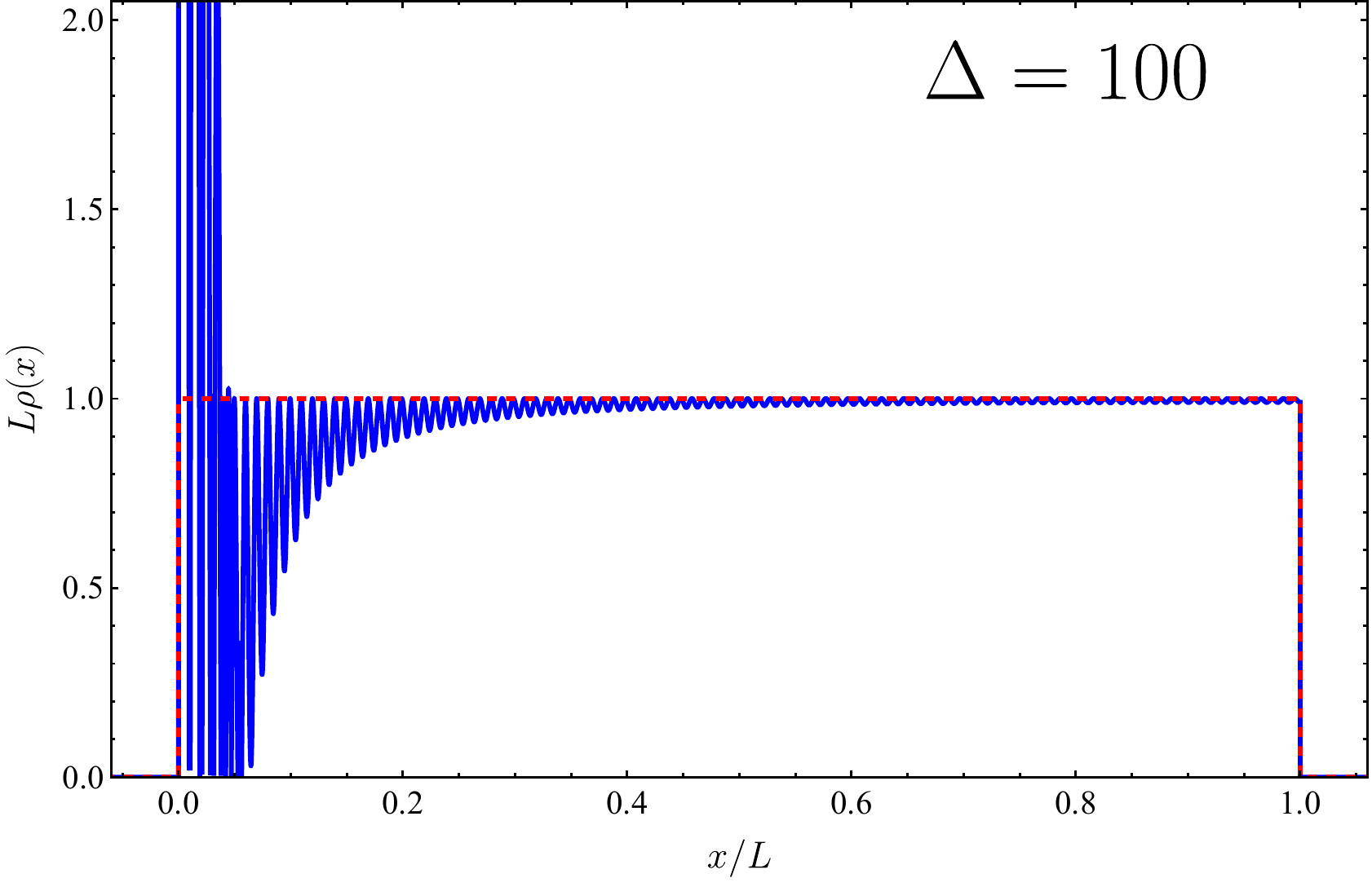} 

\caption{Plots of probability distributions for the superposition of states defined in Eq. \eqref{PD_Final} (blue) for $\Delta = 10$ (top), $\Delta = 40$ (middle) and $\Delta = 100$ (bottom). The corresponding CPD is shown for comparison (red dashed). The deviation between the quantum and classical distributions, more pronounced at lower $\Delta$ values, vanishes in the formal limit $\Delta \to \infty$.}\label{fig:AsympPD}
\end{figure}

Figure \ref{fig:AsympPD} displays the probability density obtained from Eq.~\eqref{PD_Final}
for several values of the parameter $\Delta$.
For small values of $\Delta$, the distribution exhibits pronounced oscillations and
significant deviations from the classical prediction, in agreement with the observations
reported by Cabrera and Kiwi \cite{PhysRevA.36.2995}.

As $\Delta$ increases, however, the probability density rapidly approaches the uniform
classical distribution.
In the formal limit $\Delta\to\infty$, which corresponds to a macroscopic regime where the
average energy is very large while the relative uncertainty of the energy measurement
remains small, the second term in Eq.~\eqref{PD_Final} tends to $-2$, while the last one vanishes, and the probability
density converges exactly to the classical result $\rho_{\text{cl}}(x)=1/L$.

An important feature visible in the plots is the presence of rapidly oscillatory
structures localized near the boundary at $x=0$.
This asymmetric boundary layer originates from the phase alignment of the superposition. Near $x=0$, all eigenstates $\psi_m \propto \sin(m\pi x / L)$ share a positive slope and interfere constructively. Conversely, near $x=L$, their alternating parity causes destructive interference among neighboring states, anchoring the macroscopic structure entirely to the left wall, yielding the factor
$\csc^{2}(\pi x/2L)$ appearing in Eq.~\eqref{PD_Final}. Furthermore, this boundary behavior is not a mere artifact of the rigid spatial boundaries. For instance, the graphs presented by Cabrera and Kiwi for the harmonic oscillator \cite{PhysRevA.36.2995} also exhibit an asymmetric trend despite the underlying potential being perfectly smooth. This reinforces that such extended boundary behavior is a direct consequence of the wave packet's specific superposition rather than the potential's discontinuity.

In the macroscopic limit this oscillatory boundary layer becomes confined to an
extremely narrow region of size $\Delta x \sim L/\Delta$.
Outside this vanishingly small interval the probability density becomes essentially
indistinguishable from the uniform classical distribution.
Thus, the residual quantum structure survives only at scales that are far below any
physically resolvable resolution, while the interior of the well reproduces the
classical behavior with high accuracy.

We emphasize that the choice of an equiprobable superposition in Eq.~\eqref{wave_Superp} is primarily motivated by analytical tractability, as it allows for an exact evaluation of all the relevant sums. However, one should expect that the emergence of the classical limit does not rely on this specific uniform shape. To illustrate this, let us consider a Gaussian wave packet, which serves as a standard representation of macroscopic measurement uncertainty. The state can be written as
\begin{equation}
    \Psi(x) = \mathcal{N} _{n} \sum_{k} e^{-\frac{(k-n)^2}{4\sigma^2}} \psi_{k}(x),
    \label{Eq:Gauss_WP}
\end{equation}
where $\mathcal{N} _{n}$ is a normalization constant. The parameter $\sigma$ dictates the spread in the quantum number space and acts as an analog of $\Delta$ for this distribution. In line with the above results, we operate in the macroscopic regime where this spread is large compared to the energy level spacing but small compared to the peak energy ($1 \ll \sigma \ll n$). Therefore, the corresponding probability density can be written as
\begin{equation}
    \rho(x) = \frac{H(x)H(L-x)}{L} \left| 1 + 2 \sum_{\alpha=1}^{\infty} e^{-\frac{\alpha^2}{8\sigma^2}} \cos\left( \frac{\pi \alpha x}{L} \right) \right|.
\end{equation}
This result exhibits the same qualitative behavior as the one shown in Fig.~\ref{fig:AsympPD}, i.e., it shows a boundary layer concentrated at $x=0$.

\section{Dynamical correspondence in the superposed state}
\label{Sec:Time}

Having established the static properties of the superposed state, we now turn to its dynamical behavior. Our aim is to determine whether the expectation value of the position operator reproduces the classical motion in the macroscopic regime. To this end, we consider
\begin{equation}
    \big\langle x \big\rangle (t) = \int_{0}^{L} \Psi^{\ast}(x,t) \, \hat{x} \, \Psi(x,t)\, dx,
    \label{Expectation_Def}
\end{equation}
where the time-dependent state is obtained by evolving the superposition defined in Eq.~\eqref{wave_Superp}:
\begin{equation}
\Psi(x,t) = \frac{1}{\sqrt{2\Delta +1}}\sum_{k=-\Delta}^{\Delta} e^{-i E_{n+k} t / \hbar}  \psi_{n+k}(x),
\label{wave_Superp-time}
\end{equation}
Here $n\gg1$ ensures that the state remains in the high-energy regime relevant for the correspondence limit.

Substituting Eq.~\eqref{wave_Superp-time} into Eq.~\eqref{Expectation_Def} generates diagonal and off-diagonal contributions. The diagonal part is time independent and yields the classical contribution associated with the uniform density, namely
\begin{equation}
\left\langle x \right\rangle_{\mathrm{cl}}=\frac{L}{2}.
\end{equation}
The nontrivial dynamics arises from the off-diagonal interference terms, which carry the time-dependent phases. Using the asymptotic form of the interference contributions, Eq.~\eqref{Asymp_NDTerm_inf}, together with the high-energy approximation for the level spacings, Eq.~\eqref{Diff_energies}, the contribution associated with a fixed index separation $\alpha$ can be written as
\begin{align}
    \big\langle x \big\rangle_{\alpha} (t) = &  \left( e^{-2i \frac{E_{1}}{\hbar} n \alpha t} + e^{2i \frac{E_{1}}{\hbar} n \alpha t} \right) \int_{0}^{L} x \, \rho^{\text{a}}_{\alpha}(x) dx \notag\\[6pt]
    = & \frac{4 \left[ (-1) ^{\alpha} - 1 \right] L}{\pi ^2 \alpha ^2} \cos\left(2 \frac{E_{1}}{\hbar} n \alpha t\right),
\end{align}
where $E_{1}$ denotes the ground-state energy.

Summing over all separations $\alpha$, the full expectation value becomes
\begin{equation}
    \big\langle x \big\rangle (t) = \left\langle x \right\rangle_{\mathrm{cl}} + \frac{1}{2\Delta +1} \sum_{\alpha =1}^{2\Delta} (2\Delta - \alpha +1) \big\langle x \big\rangle_{\alpha} (t).
    \label{Expect_valuet}
\end{equation}
Although the summation in the second term can also be expressed in closed form by using Eq.~\eqref{sum_st} with $N = 2\Delta$ and $\omega = 2E_1 n t / \hbar$, its explicit extended expression is not especially illuminating. For this reason, we omit the intermediate algebra in the main text; however, to generate the exact dynamical plots presented in Fig. 4, we evaluate $\langle x \rangle(t)$ using this closed-form expression from Eq.~\eqref{sum_st}.

In order to perform the numerical evaluation and connect the quantum description with its classical counterpart, we first relate the quantum parameters of the model to the classical velocity of the particle. From Eq.~\eqref{WaveFunct}, the quantized momentum is $p_n = n\pi\hbar/L$. Identifying this quantity with the classical momentum $p_n = m v_{\mathrm{cl}}$, where $v_{\mathrm{cl}}$ denotes the classical particle speed, yields the relation
\begin{equation}
n = \frac{m L v_{\mathrm{cl}}}{\pi \hbar}.
\end{equation}
Using this identification, the frequency parameter $\omega$ appearing in the sum $S_t$ can be written as
\begin{align}
    \omega = \pi \frac{v_{\mathrm{cl}}}{L} t .
\end{align}
This relation allows the quantum expectation value and the classical trajectory to be expressed in terms of the same physical parameter $v_{\mathrm{cl}}$, enabling a direct comparison between both descriptions. The motion of a classical particle bouncing elastically between the rigid walls of the box is described by a periodic triangular trajectory. Assuming the initial condition $x(0)=L$, the classical position is given by
\begin{align}
    x(t) = \big|\, (v_{\mathrm{cl}} t \bmod 2L) - L \,\big|,
    \label{Class_pos}
\end{align}
With $v_{\mathrm{cl}}$ fixed, Eqs.~\eqref{Expect_valuet} and \eqref{Class_pos} provide a direct framework for comparing the quantum and classical dynamics.

\begin{figure}
\vspace{.4cm}
\includegraphics[scale=0.23]{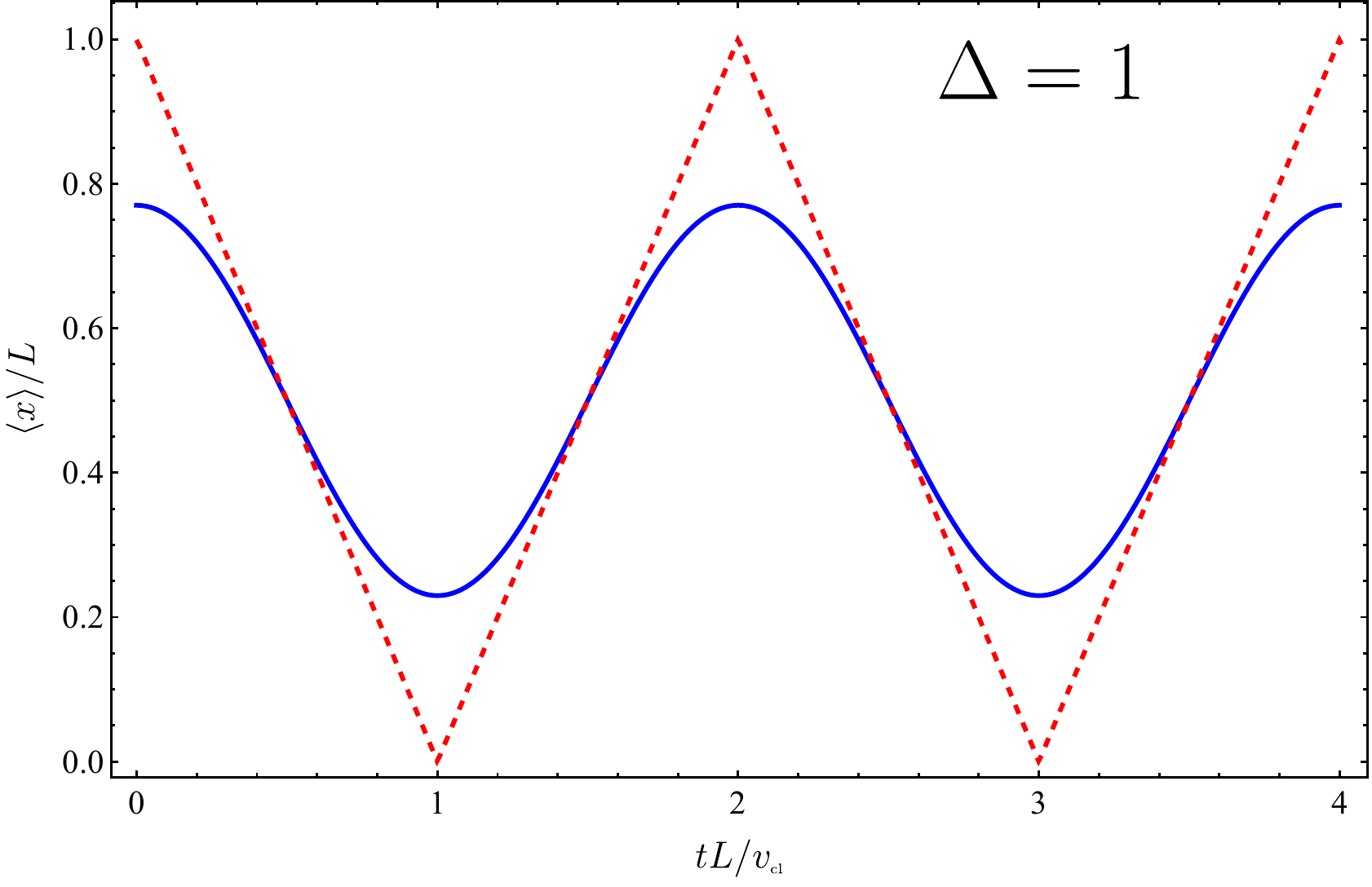}
\includegraphics[scale=0.23]{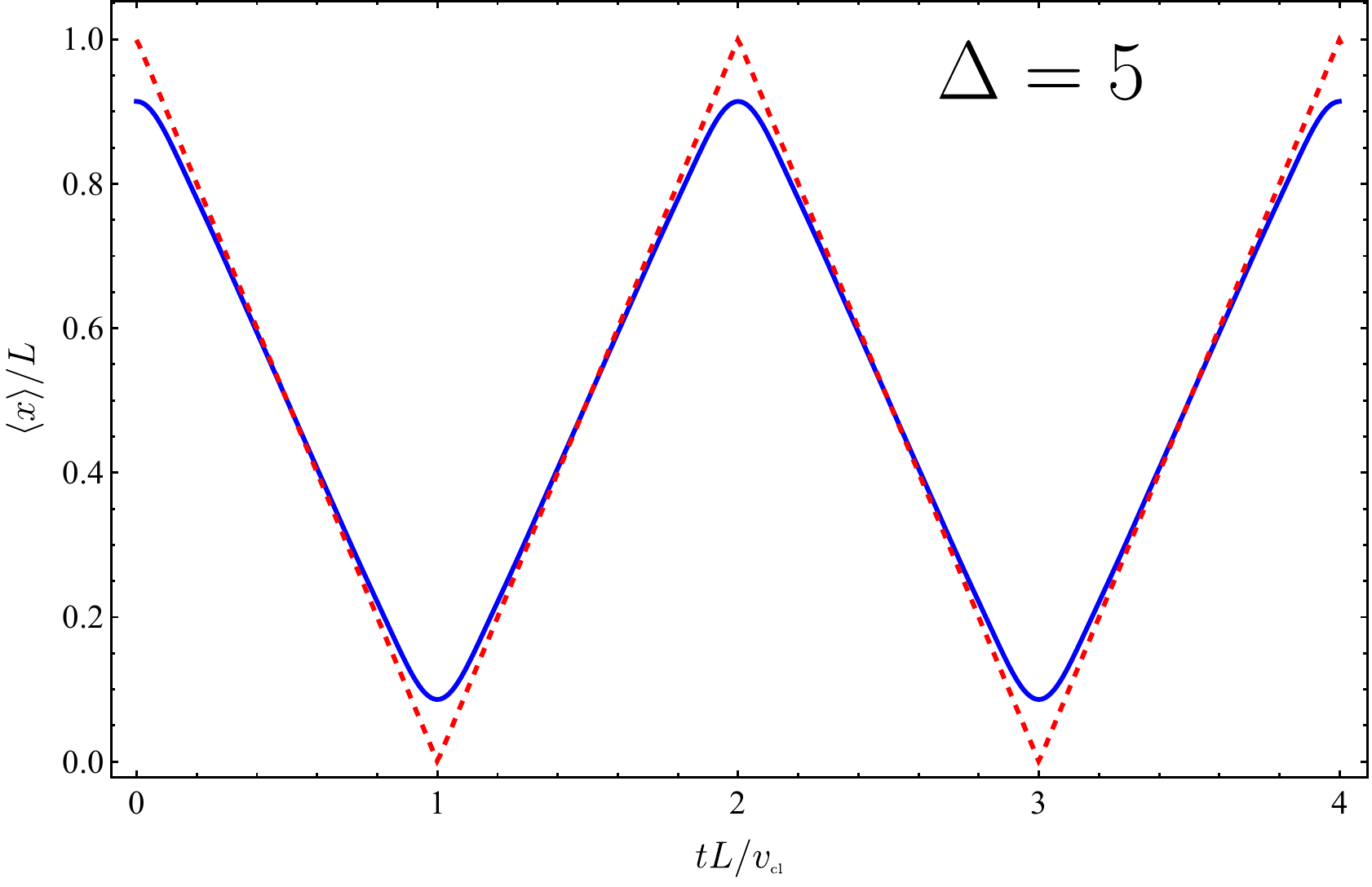}
\includegraphics[scale=0.23]{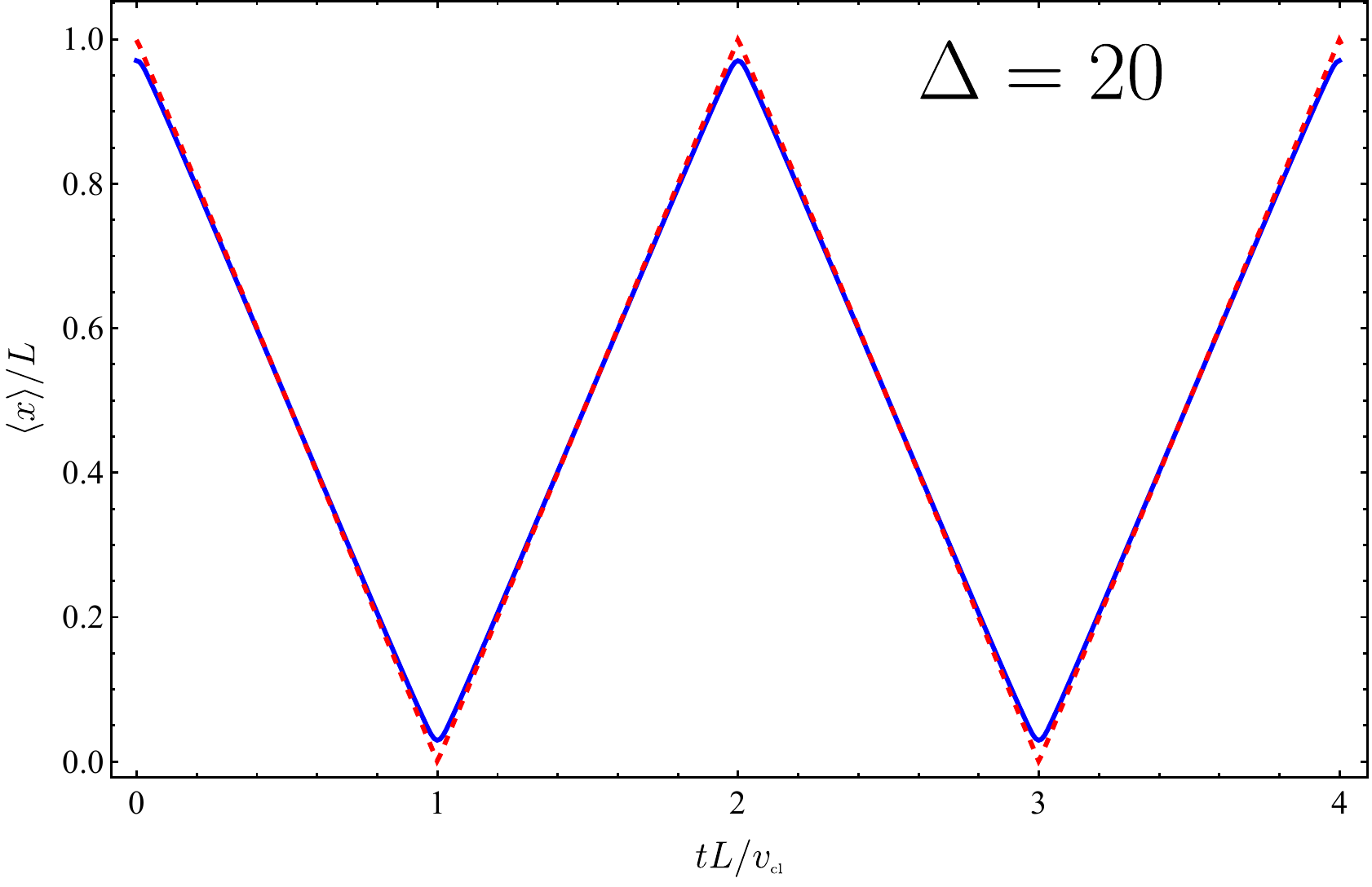} 
\caption{Plots of the time evolution of the position expectation value $\langle x \rangle$ (blue solid line) for $\Delta = 1$ (top), $\Delta = 5$ (middle) and $\Delta = 20$ (bottom). The classical trajectory $x(t)$ of a particle in the ISW is shown for comparison (red dashed line). The superposition of high-energy eigenstates clearly reproduces the classical behavior, with the agreement improving as $\Delta$ increases.}\label{fig:Position}
\end{figure}

Figure~\ref{fig:Position} shows that the high-energy expectation value $\langle x\rangle(t)$ closely follows the classical trajectory $x(t)$ over the entire period of motion. The most noticeable deviations occur near the turning points at the boundaries, where $\langle x\rangle(t)$ exhibits a slightly anticipatory behavior, as if the quantum wave packet begins to reverse its motion marginally before the classical particle reaches the wall. Such behavior is consistent with the wave-packet distortions previously reported in Ref.~\cite{10.1119/1.18854} and can be interpreted as a residual manifestation of quantum coherence in the vicinity of the boundaries.

Importantly, as the parameter $\Delta$ increases the temporal region where this distortion occurs becomes progressively narrower, confining the deviation to a small neighborhood of the turning points while leaving the bulk motion essentially classical. Moreover, a simple time translation $t\rightarrow t+t_0$ aligns the initial conditions of both curves, showing that the quantum expectation value reproduces the classical trajectory up to a global phase shift in time.

Finally, it is worth noting that the same dynamical correspondence is obtained for the Gaussian wave packet in Eq.~\eqref{Eq:Gauss_WP}. 
In this case, the expectation value of position is
\begin{equation}
     \langle x \rangle(t) = \langle x \rangle_{\text{cl}} - \frac{4L}{\pi^2} \sum_{k=1}^{\infty} \frac{1}{(2k-1)^2} e^{-\frac{(2k-1)^2}{8\sigma^2}} \cos\left( \omega (2k-1) \right).
\end{equation}
A numerical evaluation of this expression shows that it reproduces the same classical triangular trajectory found for the equiprobable distribution of states. 
For this reason, we do not include additional plots, which would be redundant.

\section{Conclusions}
\label{Sec:Conclusion}

In this work we have extended the analytical framework developed in 
Refs.~\cite{Bernal_2013,Mart_n_Ruiz_2013,Ca_as_2022, universe10090351, CMB_IJMP_2024}
for the correspondence principle, previously applied to single eigenstates of
bound systems, to the more general and physically relevant case of
superpositions of eigenstates in the infinite square well.
This extension addresses a long-standing question concerning the validity of
the correspondence principle for non-stationary states.

Our analysis is based on a Fourier-space formulation that allows the
asymptotic classical behavior of quantum probability densities to be extracted
directly from their quantum counterparts.
In contrast to approaches that invoke environmental decoherence
\cite{Zurek1991,RevModPhys.75.715, e24111520}, or those requiring thermalization and external amplification to suppress non-classical states\cite{TIRANDAZ20191677}, the present method operates entirely within configuration
space, without relying on phase-space quasi-probability distributions such as the
Wigner function. 
This makes it particularly suitable for studying the classical limit of
isolated bound systems.

By deriving a closed-form asymptotic expression for the interference
contributions $\rho_{\alpha}^{a}(x)$, obtained through a geometric expansion
of the quantum Fourier coefficients, we have shown that interference terms do
not vanish individually in the high-energy limit.
Instead, they organize into a smooth functional envelope that modulates the
rapid quantum oscillations.
When the superposition includes a macroscopic number of neighboring states,
reflecting realistic uncertainties in energy measurements at high energies,
these contributions combine in such a way that the total probability density
converges exactly to the classical uniform distribution inside the well.

Our results therefore resolve the criticism raised by Cabrera and Kiwi
\cite{PhysRevA.36.2995}, who argued that the persistence of interference terms
implies that the correspondence principle retains only heuristic value for
superpositions.
The present analysis demonstrates that the persistence of these terms does not
invalidate the correspondence principle.
Rather, their collective structure ensures that classical behavior emerges
once the macroscopic limit and realistic measurement resolutions are properly
taken into account.

The dynamical analysis further reinforces this conclusion.
The expectation value of the position operator was shown to reproduce the
classical triangular trajectory of a particle bouncing between rigid walls.
Small anticipatory distortions appear near the boundaries as a consequence of
wave-packet deformation \cite{10.1119/1.18854}, but these effects become
progressively confined to extremely narrow boundary regions as the number of
superposed states increases.
Away from these turning points, the quantum dynamics becomes practically
indistinguishable from the classical motion.

Taken together, these results establish that the correspondence principle
retains a precise and rigorous validity for superpositions of high-energy
states.
Classical behavior does not arise from the disappearance of quantum
interference, but from the collective organization of these contributions in
the macroscopic limit.
Future work may extend this formalism to other confining potentials and
many-body systems, helping to further clarify the mechanisms through which
classical physics emerges from the underlying quantum description.

\section*{Acknowledgments}
J.A.C. gratefully acknowledges the support of SECIHTI through the program \textit{Becas Nacionales para estudios de Posgrado}, under grant number 4018746. D.A.B. was supported by the DGAPA-UNAM Posdoctoral Program. A.M.-R. acknowledges financial support by UNAM-PAPIIT project No. IG100224, UNAM-PAPIME project No. PE109226, by SECIHTI project No. CBF-2025-I-1862 and by the Marcos Moshinsky Foundation.

%------------------------------------------------------

\appendix
\section{Derivation of the sum in Sec. \ref{SEC:PD}}
\label{Ap:A}
The series can be recast by transforming the linear factor into a double sum to exploit its telescopic structure:
\begin{equation}
S = \sum_{\alpha=1}^{N} (N - \alpha + 1) \cos(\alpha y) = \sum_{k=1}^{N} \sum_{\alpha=1}^{k} \cos(\alpha y).
\end{equation}
The inner sum corresponds to a Dirichlet kernel \cite{Gonzlez-Velasco1995}. Evaluating it by multiplying and dividing by $2\sin(y/2)$ and using the identity $2\sin A \cos B = \sin(A+B) - \sin(A-B)$ yields a telescoping cancellation of all intermediate terms:
\begin{align}
\sum_{\alpha=1}^{k} \cos(\alpha y) =& \frac{1}{2\sin(y/2)} \sum_{\alpha=1}^{k} \left[ \sin\left(\left(\alpha+\tfrac{1}{2}\right)y\right)\right.\notag\\
&\quad \left.- \sin\left(\left(\alpha-\tfrac{1}{2}\right)y\right) \right] \notag \\[6pt]
=& \frac{\sin\left(\left(k+\tfrac{1}{2}\right)y\right) - \sin(y/2)}{2\sin(y/2)}.
\end{align}
Substituting this result into the expression for $S$ and separating the terms, noting that the second term is constant under the sum over $k$, we obtain
\begin{equation}
S = \sum_{k=1}^{N} \frac{\sin\left(\left(k+\tfrac{1}{2}\right)y\right)}{2\sin(y/2)} - \frac{N}{2}.
\end{equation}
Applying the telescoping method once more to the remaining series, this time using $2\sin A \sin B = \cos(A-B) - \cos(A+B)$, gives
\begin{align}
\sum_{k=1}^{N} \sin\left(\left(k+\tfrac{1}{2}\right)y\right) =& \frac{1}{2\sin(y/2)} \sum_{k=1}^{N} \left[ \cos(k y) \right.\notag\\
&\quad\left.- \cos((k+1)y) \right] \notag \\[6pt]
=& \frac{\cos(y) - \cos((N+1)y)}{2\sin(y/2)}.
\end{align}
Collecting all the parts, the central sum $S$ takes the compact form
\begin{equation}
S = \frac{\cos(y) - \cos((N+1)y)}{4\sin^2(y/2)} - \frac{N}{2}.
\label{ap:sum}
\end{equation}

\section{derivation of the sum in sec. \ref{Sec:Time}}

To evaluate the summation
\begin{equation}
    S_{t} = \sum_{\alpha =1}^{N} (N - \alpha +1) \frac{(-1)^{\alpha} - 1}{\alpha^2} \cos(\omega \alpha)
\end{equation}
in closed form, we set $z = e^{i\omega}$, so that $\cos(\omega \alpha) = \frac{1}{2}(z^\alpha + z^{-\alpha})$. The alternating factor $(-1)^\alpha - 1$ implies that the sum can be constructed from a combination of the same generating structure evaluated at $z$ and $-z$. We introduce the generating function:
\begin{equation}
    F(\xi) = \sum_{\alpha=1}^{N} (N - \alpha + 1) \frac{\xi^\alpha}{\alpha^2},
\end{equation}
where the full summation is proportional to the linear combination $F(-z) - F(z) + F(-z^{-1}) - F(z^{-1})$, thus isolating the contribution of the alternating term. Distributing the linear factor $(N - \alpha + 1)$, the generating function separates into two finite series:
\begin{equation}
    F(\xi) = (N+1) \sum_{\alpha=1}^{N} \frac{\xi^\alpha}{\alpha^2} - \sum_{\alpha=1}^{N} \frac{\xi^\alpha}{\alpha}.
\end{equation}
These finite sums can be identified as truncated polylogarithmic series. They may be expressed as the difference between the corresponding infinite polylogarithm \cite{Bateman1953,Lewin1981}, $\text{Li}_s(\xi) = \sum_{\alpha=1}^{\infty} \xi^\alpha / \alpha^s$, and a remainder term accounting for the truncation at $\alpha = N$. This remainder is naturally written in terms of the Lerch transcendent \cite{Bateman1953}, $\Phi(\xi, s, a) = \sum_{k=0}^{\infty} \xi^k / (k+a)^s$. By setting the shift parameter to $a = N+1$, the finite sums become:
\begin{equation}
    \sum_{\alpha=1}^{N} \frac{\xi^\alpha}{\alpha^s} = \text{Li}_s(\xi) - \xi^{N+1} \Phi(\xi, s, N+1).
\end{equation}
Substituting this identity for $s=1$ and $s=2$ into the  expression for $F(\xi)$ yields the closed form:
\begin{align}
    F(\xi) = & (N+1) \left[ \text{Li}_2(\xi) - \xi^{N+1} \Phi(\xi, 2, N+1) \right] \notag\\[6pt]
    & - \left[ \text{Li}_1(\xi) - \xi^{N+1} \Phi(\xi, 1, N+1) \right],
\end{align}
which provides an exact representation of the generating function in terms of standard special functions. The contributions containing $\text{Li}_1(\xi)$ can be simplified by applying the identity $\text{Li}_1(\xi) = -\ln(1-\xi)$. Upon forming the full linear combination $F(-z) - F(z) + F(-z^{-1}) - F(z^{-1})$, the logarithmic contributions reorganize into differences that can be expressed in terms of inverse hyperbolic functions. In particular, they reduce naturally to combinations of $\tanh^{-1}(z)$ and $\coth^{-1}(z)$ after straightforward algebraic manipulation. The final closed-form expression is obtained by substituting $z = e^{i\omega}$ back into the expanded combination and algebraically simplifying the resulting terms, leading to
\begin{align}
S_t
=&\, \frac{e^{-i(N\omega+\omega)}}{2N^2}
\Bigg\{
e^{i\omega}
\Bigg[
2N^2 e^{iN\omega}
\Bigl(
\tanh^{-1}\!\left(e^{i\omega}\right)
\notag\\
&+\coth^{-1}\!\left(e^{i\omega}\right)
\Bigr)
+e^{2iN\omega}+1
\Bigg]
\notag\\
&
+N^2
\Bigg[
e^{i\omega}\Phi\!\left(-e^{-i\omega},1,N\right)
\notag\\
&-(N+1)e^{i\omega}\Phi\!\left(-e^{-i\omega},2,N\right)
\notag\\
&
-\Phi\!\left(e^{-i\omega},1,N+1\right)
\notag\\
&+(N+1)\Phi\!\left(e^{-i\omega},2,N+1\right)
\notag\\
&
+e^{i(2N\omega+\omega)}
\Bigg(
\Phi\!\left(-e^{i\omega},1,N\right)
\notag\\
&-(N+1)\Phi\!\left(-e^{i\omega},2,N\right)
\notag\\
&
+e^{i\omega}
\Bigl(
(N+1)\Phi\!\left(e^{i\omega},2,N+1\right)
\notag\\
&-\Phi\!\left(e^{i\omega},1,N+1\right)
\Bigr)
\Bigg)
\notag\\
&
+(N+1)e^{i(N\omega+\omega)}
\Bigg(
\operatorname{Li}_2\!\left(-e^{-i\omega}\right)
\notag\\
&
-\operatorname{Li}_2\!\left(e^{-i\omega}\right)+\operatorname{Li}_2\!\left(-e^{i\omega}\right)
-\operatorname{Li}_2\!\left(e^{i\omega}\right)
\Bigg)
\Bigg]
\Bigg\}.\label{sum_st}
\end{align}

To our knowledge, this exact representation does not provide direct analytical insight into the asymptotic regime $N \gg 1$ beyond what can be extracted numerically.

\bibliographystyle{elsarticle-num}

\bibliography{cas-refs}

@article{Makowski_2006,
    doi = {10.1088/0143-0807/27/5/012},
    url = {http://dx.doi.org/10.1088/0143-0807/27/5/012},
    issn = {1361-6404},
    year = {2006},
    month = {jul},
    pages = {1133–1139},
    title = {A brief survey of various formulations of the correspondence principle},
    number = {5},
    volume = {27},
    journal = {European Journal of Physics},
    publisher = {IOP Publishing},
    author = {Makowski, Adam J}
}

@article{Liboff_1984,
    doi = {10.1063/1.2916084},
    url = {http://dx.doi.org/10.1063/1.2916084},
    issn = {1945-0699},
    year = {1984},
    month = {feb},
    pages = {50–55},
    title = {The correspondence principle revisited},
    number = {2},
    volume = {37},
    journal = {Physics Today},
    publisher = {AIP Publishing},
    author = {Liboff, Richard L.}
}

@article{Robinett_1995,
    doi = {10.1119/1.17807},
    url = {http://dx.doi.org/10.1119/1.17807},
    issn = {1943-2909},
    year = {1995},
    month = {sep},
    pages = {823–832},
    title = {Quantum and classical probability distributions for position and momentum},
    number = {9},
    volume = {63},
    journal = {American Journal of Physics},
    publisher = {American Association of Physics Teachers (AAPT)},
    author = {Robinett, R. W.}
}

@article{Bernal_2013,
    doi = {10.4236/jmp.2013.41017},
    url = {http://dx.doi.org/10.4236/jmp.2013.41017},
    issn = {2153-120X},
    year = {2013},
    pages = {108–112},
    title = {A Simple Mathematical Formulation of the Correspondence Principle},
    number = {01},
    volume = {04},
    journal = {Journal of Modern Physics},
    publisher = {Scientific Research Publishing, Inc.},
    author = {Bernal, J. and Martín-Ruiz, A. and García-Melgarejo, J. C.}
}

@article{Mart_n_Ruiz_2013,
    doi = {10.4236/jmp.2013.46112},
    url = {http://dx.doi.org/10.4236/jmp.2013.46112},
    issn = {2153-120X},
    year = {2013},
    pages = {818–822},
    title = {The Classical Limit of the Quantum Kepler Problem},
    number = {06},
    volume = {04},
    journal = {Journal of Modern Physics},
    publisher = {Scientific Research Publishing, Inc.},
    author = {Martín-Ruiz, A. and Bernal, J. and Frank, Alejandro and Carbajal-Dominguez, Adrián}
}

@article{Ca_as_2022,
    doi = {10.1140/epjp/s13360-022-03529-2},
    url = {http://dx.doi.org/10.1140/epjp/s13360-022-03529-2},
    issn = {2190-5444},
    year = {2022},
    month = {dec},
    title = {Exact classical limit of the quantum bouncer},
    number = {12},
    volume = {137},
    journal = {The European Physical Journal Plus},
    publisher = {Springer Science and Business Media LLC},
    author = {Cañas, Juan A. and Bernal, J. and Martín-Ruiz, A.}
}

@article{Doncheski_2000,
    doi = {10.1088/0143-0807/21/3/303},
    url = {http://dx.doi.org/10.1088/0143-0807/21/3/303},
    issn = {1361-6404},
    year = {2000},
    month = {apr},
    pages = {217–228},
    title = {Comparing classical and quantum probability distributions for an asymmetric infinite well},
    number = {3},
    volume = {21},
    journal = {European Journal of Physics},
    publisher = {IOP Publishing},
    author = {Doncheski, M A and Robinett, R W}
}

@article{Robinett_1996,
    doi = {10.1119/1.18188},
    url = {http://dx.doi.org/10.1119/1.18188},
    issn = {1943-2909},
    year = {1996},
    month = {apr},
    pages = {440–446},
    title = {Visualizing the solutions for the circular infinite well in quantum and classical mechanics},
    number = {4},
    volume = {64},
    journal = {American Journal of Physics},
    publisher = {American Association of Physics Teachers (AAPT)},
    author = {Robinett, R. W.}
}

@article{Robinett_2002,
    doi = {10.1088/0143-0807/23/2/310},
    url = {http://dx.doi.org/10.1088/0143-0807/23/2/310},
    issn = {1361-6404},
    year = {2002},
    month = {jan},
    pages = {165–174},
    title = {Visualizing classical and quantum probability densities for momentum using variations on familiar one-dimensional potentials},
    number = {2},
    volume = {23},
    journal = {European Journal of Physics},
    publisher = {IOP Publishing},
    author = {Robinett, R W}
}

@article{Yoder2006UsingCP,
    url = {https://api.semanticscholar.org/CorpusID:122279845},
    year = {2006},
    pages = {404-411},
    title = {Using classical probability functions to illuminate the relation between classical and quantum physics},
    volume = {74},
    journal = {American Journal of Physics},
    author = {Garett Wade Yoder}
}

@article{Hernandez_Aguilar-Gutierrez_Bernalc_2023, place={UK}, title={On the correspondence principle for the Klein-Gordon and Dirac Equations}, volume={16}, url={https://oiccpress.com/jtap/article/view/1948}, DOI={10.30495/JTAP.162244}, abstractNote={We investigate the asymptotic behavior of the solutions to the Klein-Gordon and Dirac equations using the local spatial averaging approach to Bohr’s correspondence principle in the large principal quantum number regime. The procedure is applied in two basic problems in $1+1$-dimensions, the relativistic quantum oscillator and the relativistic particle in a box. In the harmonic oscillator cases, we find that the corresponding probability densities reduce to their respective classical single-particle distributions plus a series of terms suppressed by powers of the $hbar$ constant, while particle in a box cases show a different structure for the quantum corrections.}, number={4}, journal={Journal of Theoretical and Applied Physics}, author={Hernandez, Kevin G. and Aguilar-Gutierrez, Sergio E. and Bernalc, Jorge}, year={2023}, month={Nov.} }

@Article{universe10090351,
AUTHOR = {Cañas, Juan A. and Bernal, J. and Martín-Ruiz, A.},
TITLE = {On the Classical Limit of Freely Falling Quantum Particles, Quantum Corrections and the Emergence of the Equivalence Principle},
JOURNAL = {Universe},
VOLUME = {10},
YEAR = {2024},
NUMBER = {9},
ARTICLE-NUMBER = {351},
URL = {https://www.mdpi.com/2218-1997/10/9/351},
ISSN = {2218-1997},
DOI = {10.3390/universe10090351}
}

@article{CMB_IJMP_2024,
author = {Ca\~{n}as, Juan A. and Mart\'{\i}n-Ruiz, A. and Bernal, J.},
title = {Emergent universality of free fall from quantum mechanics},
journal = {International Journal of Modern Physics D},
volume = {33},
number = {15},
pages = {2441004},
year = {2024},
doi = {10.1142/S0218271824410049},
URL = {https://doi.org/10.1142/S0218271824410049},
eprint = {https://doi.org/10.1142/S0218271824410049}
}

@article{PhysRevA.36.2995,
  title = {Large quantum-number states and the correspondence principle},
  author = {Cabrera, G. G. and Kiwi, Miguel},
  journal = {Phys. Rev. A},
  volume = {36},
  issue = {6},
  pages = {2995--2998},
  numpages = {0},
  year = {1987},
  month = {Sep},
  publisher = {American Physical Society},
  doi = {10.1103/PhysRevA.36.2995},
  url = {https://link.aps.org/doi/10.1103/PhysRevA.36.2995}
}

@book{Gradshteyn2015,
   author = {I.S. Gradshteyn and I.M. Ryzhik},
   city = {Boston},
   doi = {10.1016/C2010-0-64839-5},
   edition = {8},
   isbn = {9780123849335},
   publisher = {Academic Press},
   title = {Table of Integrals, Series, and Products},
   year = {2015}
}

@book{Neumann2018,
   author = {John von. Neumann and Nicholas A.. Wheeler},
   isbn = {9780691178561},
   publisher = {Princeton University Press : Princeton University Press},
   title = {Mathematical foundations of quantum mechanics},
   year = {2018}
}

@article{Zurek1991,
   author = {Wojciech H. Zurek},
   doi = {10.1063/1.881293},
   issn = {0031-9228},
   issue = {10},
   journal = {Physics Today},
   month = {10},
   pages = {36-44},
   title = {Decoherence and the Transition from Quantum to Classical},
   volume = {44},
   year = {1991}
}

@article{Rosen1964,
   author = {Nathan Rosen},
   doi = {10.1119/1.1970870},
   issn = {0002-9505},
   issue = {8},
   journal = {American Journal of Physics},
   month = {8},
   pages = {597-600},
   title = {The Relation Between Classical and Quantum Mechanics},
   volume = {32},
   year = {1964}
}

@article{Home1984,
   author = {D. Home and S. Sengupta},
   doi = {10.1007/BF02732874},
   issn = {1826-9877},
   issue = {2},
   journal = {Il Nuovo Cimento B Series 11},
   month = {8},
   pages = {214-224},
   title = {Classical limit of quantum mechanics. A paradoxical example},
   volume = {82},
   year = {1984}
}

@article{Korsch1978,
   author = {H.J Korsch and R Möhlenkamp},
   doi = {10.1016/0375-9601(78)90035-X},
   issn = {03759601},
   issue = {2},
   journal = {Physics Letters A},
   month = {7},
   pages = {110-112},
   title = {A note on multidimensional WKB wavefunctions: Local and global semiclassical approximations},
   volume = {67},
   year = {1978}
}

@article{Kmmel1955,
   author = {H. K\"ummel},
   doi = {10.1007/BF02731413},
   issn = {0029-6341},
   issue = {6},
   journal = {Il Nuovo Cimento},
   month = {6},
   pages = {1057-1078},
   title = {Zur quantentheoretischen Begr\"undung der klassischen Physik},
   volume = {1},
   year = {1955}
}

@incollection{Gonzlez-Velasco1995,
  author    = {Enrique A. Gonz{\'a}lez-Velasco},
  title     = {Fourier Series},
  booktitle = {Fourier Analysis and Boundary Value Problems},
  pages     = {23--83},
  publisher = {Elsevier},
  year      = {1995},
  doi       = {10.1016/B978-012289640-8/50002-2}
}

@book{Bateman1953,
   author = {Harry Bateman and Arthur Erde\'elyi},
   isbn = {9780070195455},
   pages = {302},
   publisher = {McGraw-Hill},
   title = {Higher transcendental functions},
   year = {1953}
}

@book{Lewin1981,
   author = {Leonard Lewin},
   isbn = {0444005501},
   pages = {359},
   publisher = {North Holland},
   title = {Polylogarithms and associated functions},
   year = {1981}
}

@article{10.1119/1.18854,
    author = {Andrews, Mark},
    title = {Wave packets bouncing off walls},
    journal = {American Journal of Physics},
    volume = {66},
    number = {3},
    pages = {252-254},
    year = {1998},
    month = {03},
    issn = {0002-9505},
    doi = {10.1119/1.18854},
    url = {https://doi.org/10.1119/1.18854},
}

@article{RevModPhys.75.715,
  title = {Decoherence, einselection, and the quantum origins of the classical},
  author = {Zurek, Wojciech Hubert},
  journal = {Rev. Mod. Phys.},
  volume = {75},
  issue = {3},
  pages = {715--775},
  numpages = {0},
  year = {2003},
  month = {May},
  publisher = {American Physical Society},
  doi = {10.1103/RevModPhys.75.715},
  url = {https://link.aps.org/doi/10.1103/RevModPhys.75.715}
}

@Article{e24111520,
AUTHOR = {Zurek, Wojciech Hubert},
TITLE = {Quantum Theory of the Classical: Einselection, Envariance, Quantum Darwinism and Extantons},
JOURNAL = {Entropy},
VOLUME = {24},
YEAR = {2022},
NUMBER = {11},
ARTICLE-NUMBER = {1520},
URL = {https://www.mdpi.com/1099-4300/24/11/1520},
PubMedID = {36359613},
ISSN = {1099-4300},
DOI = {10.3390/e24111520}
}

@article{ALBRECHT2023169289,
title = {Bouncing wave packets, Ehrenfest theorem, and uncertainty relation based upon a new concept for the momentum of a particle in a box},
journal = {Annals of Physics},
volume = {452},
pages = {169289},
year = {2023},
issn = {0003-4916},
doi = {https://doi.org/10.1016/j.aop.2023.169289},
url = {https://www.sciencedirect.com/science/article/pii/S000349162300074X},
author = {I. Albrecht and J. Herrmann and A. Mariani and U.-J. Wiese and V. Wyss}
}

@article{10.1063/5.0178419,
    author = {Mariani, A. and Wiese, U.-J.},
    title = {Self-adjoint momentum operator for a particle confined in a multi-dimensional cavity},
    journal = {Journal of Mathematical Physics},
    volume = {65},
    number = {4},
    pages = {042102},
    year = {2024},
    month = {04},
    issn = {0022-2488},
    doi = {10.1063/5.0178419},
    url = {https://doi.org/10.1063/5.0178419},
}

@article{Fein2019,
   author = {Yaakov Y. Fein and Philipp Geyer and Patrick Zwick and Filip Kiałka and Sebastian Pedalino and Marcel Mayor and Stefan Gerlich and Markus Arndt},
   doi = {10.1038/s41567-019-0663-9},
   issn = {1745-2473},
   issue = {12},
   journal = {Nature Physics},
   month = {12},
   pages = {1242-1245},
   title = {Quantum superposition of molecules beyond 25 kDa},
   volume = {15},
   year = {2019}
}

@article{
Bild2023,
author = {Marius Bild  and Matteo Fadel  and Yu Yang  and Uwe von Lüpke  and Phillip Martin  and Alessandro Bruno  and Yiwen Chu },
title = {Schrödinger cat states of a 16-microgram mechanical oscillator},
journal = {Science},
volume = {380},
number = {6642},
pages = {274-278},
year = {2023},
doi = {10.1126/science.adf7553},
URL = {https://www.science.org/doi/abs/10.1126/science.adf7553},
eprint = {https://www.science.org/doi/pdf/10.1126/science.adf7553}}

@article{BELLONI201425,
title = {The infinite well and Dirac delta function potentials as pedagogical, mathematical and physical models in quantum mechanics},
journal = {Physics Reports},
volume = {540},
number = {2},
pages = {25-122},
year = {2014},
issn = {0370-1573},
doi = {https://doi.org/10.1016/j.physrep.2014.02.005},
url = {https://www.sciencedirect.com/science/article/pii/S037015731400043X},
author = {M. Belloni and R.W. Robinett}
}

@article{10.1063/1.1504885,
    author = {Dias, Nuno Costa and Prata, João Nuno},
    title = {Wigner functions with boundaries},
    journal = {Journal of Mathematical Physics},
    volume = {43},
    number = {10},
    pages = {4602-4627},
    year = {2002},
    month = {10},
    issn = {0022-2488},
    doi = {10.1063/1.1504885},
    url = {https://doi.org/10.1063/1.1504885},
}

@article{TIRANDAZ20191677,
title = {Classicalization of quantum state of detector by amplification process},
journal = {Physics Letters A},
volume = {383},
number = {15},
pages = {1677-1682},
year = {2019},
issn = {0375-9601},
doi = {https://doi.org/10.1016/j.physleta.2019.02.039},
url = {https://www.sciencedirect.com/science/article/pii/S0375960119301860},
author = {Arash Tirandaz and Farhad {Taher Ghahramani} and Ali Asadian and Mehdi Golshani}
}

@book{Schlosshauer2007,
  author    = {Schlosshauer, Maximilian},
  title     = {Decoherence and the {Quantum-to-Classical} Transition},
  series    = {The Frontiers Collection},
  publisher = {Springer Berlin Heidelberg},
  address   = {Berlin, Heidelberg},
  year      = {2007},
  edition   = {1},
  doi       = {10.1007/978-3-540-35775-9},
  isbn      = {978-3-540-35773-5},
  eisbn     = {978-3-540-35775-9},
  url       = {https://link.springer.com/book/10.1007/978-3-540-35775-9},
  note      = {Corrected third printing, 2008}
}

@book{Joos2003,
  author    = {Joos, Erich and Zeh, H. Dieter and Kiefer, Claus and Giulini, Domenico and Kupsch, Joachim and Stamatescu, Ion-Olimpiu},
  title     = {Decoherence and the Appearance of a Classical World in Quantum Theory},
  publisher = {Springer Berlin Heidelberg},
  address   = {Berlin, Heidelberg},
  year      = {2003},
  edition   = {2},
  doi       = {10.1007/978-3-662-05328-7},
  isbn      = {978-3-540-00390-8},
  eisbn     = {978-3-662-05328-7},
  url       = {https://link.springer.com/book/10.1007/978-3-662-05328-7},
  note      = {Softcover reprint of the hardcover second edition}
}

@incollection{Hornberger2009,
  author    = {Hornberger, Klaus},
  title     = {Introduction to Decoherence Theory},
  booktitle = {Entanglement and Decoherence: Foundations and Modern Trends},
  editor    = {Buchleitner, Andreas and Viviescas, Carlos and Tiersch, Markus},
  series    = {Lecture Notes in Physics},
  volume    = {768},
  pages     = {221--276},
  publisher = {Springer Berlin Heidelberg},
  address   = {Berlin, Heidelberg},
  year      = {2009},
  doi       = {10.1007/978-3-540-88169-8_5},
  isbn      = {978-3-540-88168-1},
  eisbn     = {978-3-540-88169-8},
  url       = {https://link.springer.com/chapter/10.1007/978-3-540-88169-8_5}
}

%\bibliographystyle{unsrt}



\end{document}